\documentclass[prfluids,reprint,amsmath,amssymb]{revtex4-2}

\usepackage{graphicx} 

\usepackage[T1]{fontenc} 

\usepackage{textcomp} 
\usepackage{gensymb} 

\usepackage[english]{babel}

\begin{document}

\title{Tunable transport in bi-disperse porous materials with vascular structure}

\author{Olivier Vincent}
\email{olivier.vincent@cnrs.fr}
\altaffiliation[Present address: ]{Universite Claude Bernard Lyon 1, CNRS, Institut Lumi\`ere Mati\`ere, UMR5306, F69100 Villeurbanne, France}
\affiliation{Robert Frederick Smith School of Chemical and Biomolecular Engineering, Cornell University, Ithaca, New York 14853, USA}

\author{Th\'eo Tassin}
\affiliation{Robert Frederick Smith School of Chemical and Biomolecular Engineering, Cornell University, Ithaca, New York 14853, USA}

\author{Erik J. Huber}
\affiliation{Sibley School of Mechanical and Aerospace Engineering, Cornell University, Ithaca, New York 14853, USA}

\author{Abraham D. Stroock}
\email{ads10@cornell.edu}
\affiliation{Robert Frederick Smith School of Chemical and Biomolecular Engineering, Cornell University, Ithaca, New York 14853, USA}
\affiliation{School of Integrative Plant Science, Cornell University, Ithaca, New York 14853, USA}
\affiliation{Kavli Institute at Cornell for Nanoscale Science, Cornell University, Ithaca, New York 14853, USA}

\begin{abstract}

We study transport in synthetic, bi-disperse porous structures, with arrays of microchannels interconnected by a nanoporous layer.
These structures are inspired by the xylem tissue in vascular plants, in which sap water travels from the roots to the leaves to maintain hydration and carry micronutrients.
We experimentally evaluate transport in three conditions: high pressure-driven flow, spontaneous imbibition, and transpiration-driven flow.
The latter case resembles the situation in a living plant, where bulk liquid water is transported upwards in a metastable state (negative pressure), driven by evaporation in the leaves;
here we report stable, transpiration-driven flows down to $\sim -15$ MPa of driving force.
By varying the shape of the microchannels, we show that we can tune the rate of these transport processes in a predictable manner, using a simple analytical (effective medium) approach and numerical simulations of the flow field in the bi-disperse media.
We also show that the spontaneous imbibition behavior of a single structure -- with fixed geometry -- can behave very differently depending on its preparation (filled with air, vs. evacuated), because of a dramatic change in the conductance of vapor in the microchannels;
this change offers a second way to tune the rate of transport in bi-disperse, xylem-like structures, by switching between air-filled and evacuated states.

\end{abstract}

\maketitle

\section{Introduction}

Porous media are ubiquitous in natural -- rocks, soil, living tissues -- and technological -- separation media, wicks, building materials etc. -- contexts. In many situations, permeation of both liquids and gases through these materials defines their function or performance \cite{Sahimi1993,Huber2015,Bacchin2021}.
In many important contexts, porous media also contain pores of vastly different (effective) lengths and diameters. This diversity of structure often can occur due to
defects (e.g., in packed columns \cite{Reising2017}),
fracture (e.g., in geological formations \cite{Huber2018}),
functional design (e.g., in catalyst supports \cite{Stuecker2004}),
or adaptive evolution (e.g., vascularized tissues) \cite{Stroock2014}.

\begin{figure}[h]
	\includegraphics[scale=1]{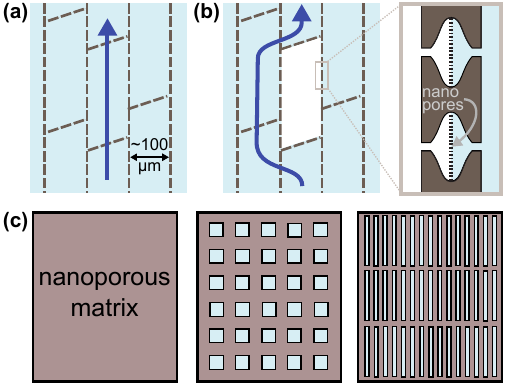}
		\caption{\small
		Xylem-inspired, bi-disperse porous media. (a-b) Schematic drawings of transport in xylem vessels before (a) and after (b) embolism. Vessels ($\sim 10-100 \, \mathrm{\mu m}$ in width) are interconnected by nanoporous membranes (pit membranes) that prevent propagation of the embolism to adjacent vessels. (c) Structures inspired by xylem and focus of this study, combining a nanoporous matrix and regular arrays of microchannels of various aspect ratios and approximately constant areal density.
		}
		\label{fig:Intro}
\end{figure}

In biology, vascular structure defines a coherent network of macroscopic vessels of high hydraulic conductivity (diameter $\sim 10-100 \, \mathrm{\mu m}$) within tissues of smaller pore size (nanometer range) of lower hydraulic conductivity \cite{Tyree2002,Nobel2020}. In the xylem of vascular plants (Figure \ref{fig:Intro}a-b), segments of macroscopic vessels are interconnected, axially and laterally, by nanoporous membranes. The xylem moves liquid water in the metastable state of negative pressure due to the undersaturation in the surrounding environment that drives the fluid motion \cite{Stroock2014}; this state is prone to the entry or emergence of vapor bubbles by cavitation or embolization. Xylem's bi-disperse, interconnected architecture is hypothesized to serve to mitigate the loss of conductance upon loss of liquid connectivity by arresting the spread of vapor and providing alternative paths for liquid flow (Figure \ref{fig:Intro}b).
This xylem structure can remain partially intact in lumber such that transport of moisture in xylem-like geometries may be important in the preparation and functionality of wood for construction and other applications. \cite{Desmarais2016,Zhou2018a,Elustondo2023}.


As we have presented in previous studies, synthetic systems inspired by the structure and hypothesized function of xylem provide a route to generating extreme scenarios in which the pore liquid experiences large tensions and to studying imbibition and permeation dynamics driven by these stresses \cite{Wheeler2008,Vincent2014,Pagay2014}.
Such \emph{synthetic xylem} structures provide a basis for testing biophysical hypotheses \cite{Wheeler2008,Vincent2019}, interrogating nanoscale transport and thermodynamic phenomena \cite{Vincent2016}, and designing approaches to heat transfer \cite{Chen2016a,Shi2020} and separations \cite{Wang2020}.

In these applications, as in vascular plants, the porous medium serves as a wick, i.e., a structure in which capillary stresses drive flow; the maximization of permeability (requiring large pores) while maintaining the ability to hold the pore liquid under tension by capillarity (requiring small pores) present competing demands on the pore structure.
The bi-disperse pore structure of xylem (Figure \ref{fig:Intro}a-b), with long, large diameter pores (vessel segments) interconnected by thin membranes of nanoporous material, would seem, intuitively, to present a favorable solution by maximizing the distance traveled in vessels of high permeability while isolating each segment from others and from the outside environment with small pores capable of generating large capillary stresses.

In this study, we build and observe permeation in a series of geometries inspired by xylem (Figure \ref{fig:Intro}c) in order to characterize the impact of pore geometry on effective permeability.
Experiments with the pore space completely filled with liquid (water) allow us to confront the predictions of both a numerical model and a simple effective medium approach of permeation in our geometries.
We extend our experiments to imbibition in which liquid advances into empty pores and elucidate a strong dependence of the rate (and thus effective permeability) on the initial presence of air or its absence (i.e., pure vapor in vacuum) in the pores. We explain this striking difference with a change in the physical mechanisms governing transport in the vascular microstructure, coupled to liquid capillary flow in the surrounding nanoporous layer.

Globally, our investigations demonstrate that bio-inspired porous systems can be fabricated to achieve tunable water transport properties that depend on both their geometries and their preparation (evacuation or not). Simple physics-based design rules allow us to predict the behavior of the structures as a function of these parameters. This type of experimental system can also be used as experimental micromodels for studying transport and phase change processes related to plants, geology, and engineered materials

\section{Methods}

\subsection{Samples}

\begin{figure*}[ht]
	\includegraphics[scale=1]{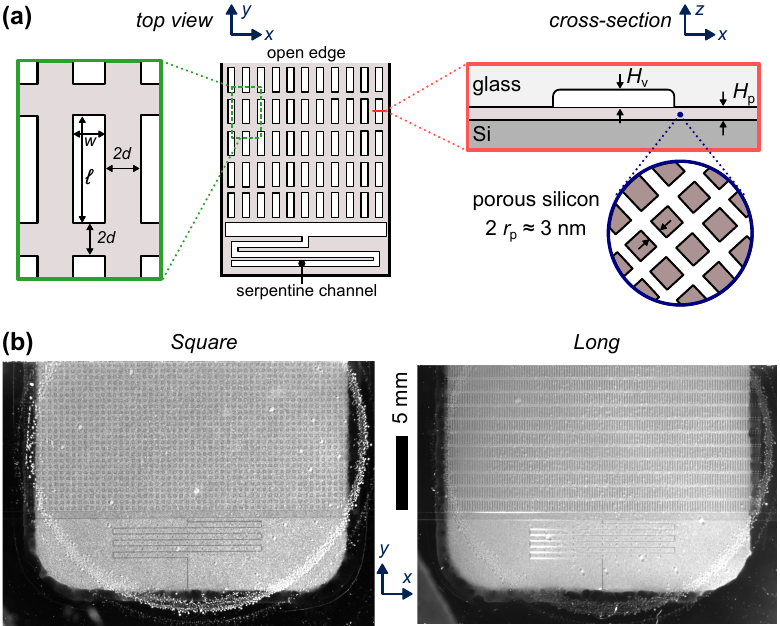}
		\caption{\small
		Synthetic, xylem-like samples studied here.
		(a) Schematic drawing: global, top view (left), and cross-section detail (right).
		In the left panel, the inset shows the definitions of the lateral dimensions of the microstructures,  $w$, $\ell$ and $d$; in the right panel,  $H_\mathrm{v}$, $H_\mathrm{p}$ and $r_\mathrm{p}$ are the microchannel depth, the nanoporous medium depth and the nanopore radius, respectively.
		See Table \ref{tab:Samples} for parameter values.
		(b) Micrographs of two designs.
		The central, gray zone is where the top surface of the silicon wafer has been anodized to create a layer of porous silicon;
		in the dark area that surrounds it, silicon is not anodized and the direct bonding of silicon to glass forms an impermeable boundary on the porous zone.
		The bonded structure is cut with a dicing saw to expose the porous silicon layer to the outside environment (top edge in micrographs);
		this cut does not open microchannels directly to the outside.
		The irregular white markings with a general circular shape are residues left by the electrode during the anodic bonding procedure;
		they are located on the external surface of the glass and did not affect the phenomena studied here.
		}
	\label{fig:Samples}
\end{figure*}

We fabricated xylem-like, composite structures with regularly spaced microchannels interconnected through a nanoporous matrix by assembling (through anodic bonding) two surfaces.
The first surface consisted of a nanoporous silicon layer (poSi) etched into the surface of a silicon wafer by anodization.
The second surface was glass, patterned with microchannels by photolithography (Figure \ref{fig:Samples}).
Details on the fabrication methods can be found elsewhere \cite{Vincent2014, Vincent2016}.
Separating the fabrication of the nanoporous layer and the microchannels in two separate surfaces results in simpler fabrication procedures compared to creating all structures into the silicon wafer
(e.g. wet etching instead of reactive ion etching, parallel instead of sequential processing etc.), but the latter is also possible \cite{Vincent2019}.
We do not expect these two fabrication procedures to result in significantly different behavior, as the final layers remain essentially two-dimensional (nanoporous layer and microchannel depths small compared to other system dimensions).


Figure \ref{fig:Samples}a presents the geometrical properties of the xylem-like designs. Figure \ref{fig:Samples}b presents micrographs of two of the geometries studied.
Table \ref{tab:Samples} provides parameter values of all designs.
In all designs, the poSi layer had a thickness, $H_\mathrm{p} = 5 \, \mathrm{\mu m}$ and contained interconnected pores of radius, $r_\mathrm{p} \simeq 1.5$ nm (as characterized previously in Vincent et al. \cite{Vincent2014,Vincent2016,Vincent2017a}).
The microchannel segments had a depth, $H_\mathrm{v} = 32 \, \mathrm{\mu m}$ (except for the \emph{Vein} sample, $H_\mathrm{v} \simeq 20 \, \mathrm{\mu m}$) and various in-plane lengths ($\ell$) and widths ($w$).
These channels were arranged in various regular, rectilinear arrays with the same edge-to-edge separation ($2d$) in both principal directions ($x$ and $y$);
the value of $d$ varied with the lateral dimensions to maintain a similar areal fraction of the microchannels across designs: $\gamma \sim 0.3$.
A sample with no microchannels (\emph{Blank}) was also prepared.

\begin{table*}[ht]
	\begin{tabular}{c | c | c | c | c | c | c | c}
		Sample  & $w \, (\mathrm{\mu m})$ & $\ell \, (\mathrm{\mu m})$ & $d \, (\mathrm{\mu m})$ &  $\xi$  & $\gamma$ & $\kappa / \kappa_0$ (simul.) & $\kappa / \kappa_0$ (approx.) \\
		\hline
		Blank   &  n.a.                   & n.a.                       & n.a                     &  n.a.   &  n.a. & 1  & 1 \\
		Square  &  224                    & 224                        & 88                      &  1      &  0.31 & 2.00 & 2.27 \\
		Short   &  124                    & 424                        & 77                      & 3.4    &  0.33  & 3.24 & 3.75 \\
		Long    &  74                     & 824                        & 50                      & 11     &  0.38  & 7.88 & 9.24 \\
		Vein    &  100                    & 14500                      & 250                     & 145    &  0.24  & 24.6 & 30.0 \\
	\end{tabular}
	\caption{
		Sample properties: dimensions of microchannel segments ($w$, $\ell$, $d$) as defined in Figure \ref{fig:Samples}, and corresponding aspect ratio ($\xi$) and area coverage ratio ($\gamma$); $\kappa / \kappa_0$ is the permeability increase predicted by our numerical simulations (see section \ref{sec:Theory}, \emph{Theory}), or by our approximate analytical formula (Equation \ref{eq:AnalyticalPermeability}), except for the \emph{Blank} sample for which $\kappa / \kappa_0 = 1$ by definition.
		The Blank sample contained no microchannels such that the geometrical parameters are not applicable (n.a.) in this case.
	}
	\label{tab:Samples}
\end{table*}

As in our previous work \cite{Vincent2016}, we also etched a serpentine channel connected to a distribution channel on the backside of the sample; we used this channel to measure the flow rate through the structure by measuring its rate of filling/emptying by tracking the progression of a meniscus.
Both of these channels were of depth $H_\mathrm{v}$, and of width $10 \, \mathrm{\mu m} + 2 H_\mathrm{v}$ (serpentine) and $460 \, \mathrm{\mu m} + 2 H_\mathrm{v}$ (serpentine)

After anodic bonding, the whole structure was closed on three sides (thick black line in Figure \ref{fig:Samples}a) and open on one edge at which mass exchange with the environment occurred from the exposed cross-section of the poSi layer; we opened this edge by cutting the anodically bonded structure formed of glass and silicon.

We fabricated five different geometries for the microstructure: a reference sample (\emph{Blank}) without any etched microchannels in the glass, serving as a reference for the measure of transport in the bare porous silicon layer (permeability $\kappa_0 \, \mathrm{[m^2 / (Pa.s)]}$), and four structures of similar surface coverage ($\gamma = S_\mathrm{microchannels} / S_\mathrm{poSi} \mathrm{[-]}$, where $S_\mathrm{microchannels} \, \mathrm{[m^2]}$ is the in-plane area of all microchannels and $S_\mathrm{poSi}$ is the total in-plane area of the porous silicon through which transport occurred), but strongly varying in-plane aspect ratio ($\xi \mathrm{[-]}$).
Table \ref{tab:Samples} provides the geometries of the features of the five samples;
see also Appendix A, Figure \ref{fig:SamplePictures} for actual micrographs of every sample.
For all samples, the structure was repeated across a zone of length $L=1$ cm and width $W=2.2$ cm, except for the \emph{vein} sample with $L = 1.5$ cm.
As a result, each vein in this latter sample spanned almost the whole length of the medium, while for other samples the pattern was repeated in the $y$ direction.

\subsection{Experiments \label{sec:MethodsExperiments}}

\begin{figure*}[ht]
	\includegraphics[scale=1]{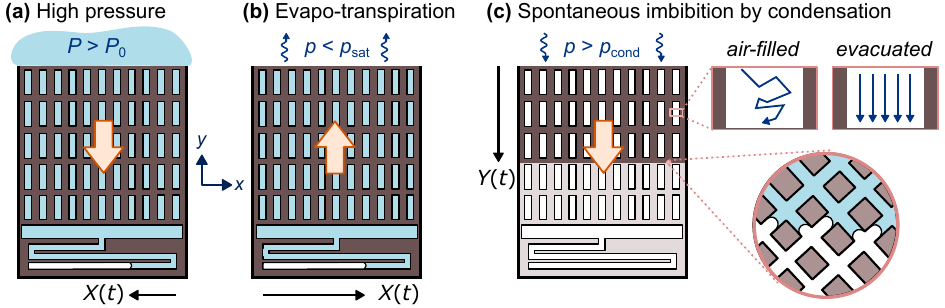}
		\caption{\small
		Different types of experiments performed on an individual sample.
		(a) High pressure flow, with measurement of flow rate using the position, $X(t)$, of the meniscus in the serpentine channel or in the distribution channel (horizontal).
		(b) Transpiration-driven emptying in a sub-saturated vapor, with a similar measurement of the flow rate using $X(t)$.
		(c) Spontaneous imbibition triggered by capillary condensation in the nanoporous matrix, with the microstructure either filled with air (diffusive transport of water vapor, represented by the blue, broken arrow), or evacuated (convective transport of water vapor, parallel blue arrows, corresponding to a Poiseuille flow developed in the $z$ direction, see Figure \ref{fig:Mechanisms}c); $Y(t)$ corresponds to the position of the wetting front, which we measure optically.
		Large, orange arrows indicate the direction of the flow, dark brown shading indicates wet nanoporous matrix, and light brown shading indicates dry nanoporous matrix.
		}
	\label{fig:Methods}
\end{figure*}

With each of the samples described above, we performed 3 distinct experiments (see Figure \ref{fig:Methods}): high pressure flow, transpiration at negative pressure, and spontaneous imbibition triggered by capillary condensation;  for imbibition, we also considered two different situations, with the microstructure either filled with air or evacuated.
All experiments were carried out at ambient lab temperature ($23 \pm 1 \degree$C), except experiments in vacuum, where the stage had a thermal regulation at $15\degree$C.
For all series of experiments, we report the observed permeability relative to that of the \emph{Blank} sample under the same conditions; we expect that this ratio should be relatively insensitive to the influence of temperature on physical parameters (e.g., viscosity).


\paragraph{High pressure flow \label{sec:MethodsPBomb}}

For high pressure flow experiments (Figure \ref{fig:Methods}a), we first removed air from the pores and microchannels by evacuation in vacuum.
Then, we placed the samples in a commercial, high-pressure vessel (HIP Inc.)
\footnote{
	Note that appropriate equipment and protection must be used when handling fluids at high pressures.
}
filled with water, with an applied pressure, $P$ (typically 3 to 20 MPa depending on the sample).
This high pressure resulted in an inwards flow, progressively filling the serpentine channel and shrinking the bubble at the backend of the sample (bottom of structures in Figure \ref{fig:Samples}).
Samples were evacuated prior to filling to avoid the pressure associated with compressing and dissolving trapped air.
We left the samples under pressure for intervals of time, $\Delta t$ before releasing the pressure.
We took snapshots of the samples out of the vessel with a stereoscope and camera (Leica MZFLIII stereoscope from Leica Microsystems GmbH, Wetzlar, Germany, and a QImaging Retiga 1300 camera from QImaging, Surrey, Canada),
in order to measure the displacement, $\Delta X$ of the meniscus in the serpentine channel (see Appendix \ref{sec:AppendixPBomb}, Figure \ref{fig:ForcedFilling});
we repeated this process until filling was complete. We estimated the velocity for each period of pressurization, $U = \Delta X / \Delta t$. This process was required because the pressure vessel was opaque.
We then estimated the mass flow rate, Q [kg/s] as:
\begin{equation}
	Q = \rho s U
	\label{eq:MassFlowRate}
\end{equation}
where $\rho \, \mathrm{[kg/m^3]}$ is the fluid's density, $s \, \mathrm{[m^2]}$ is the cross-section area of the serpentine channel, and $U \, \mathrm{[m/s]}$ is the measured velocity.

We note that in some experiments, the meniscus was in the distribution channel rather than in the serpentine channel (see e.g. Appendix \ref{sec:AppendixPBomb}, Figure \ref{fig:ForcedFilling}). In both cases, we used the known cross-sectional area of the channel, $s \, \mathrm{[m^2]}$, in which we tracked the meniscus to convert the speed of the meniscus into a volumetric flow rate.

\paragraph{Transpiration}

In this scenario, inspired by xylem sap flow in plants, flow through the structure was driven by evaporation at the open edge (the "leaf") into a subsaturated vapor (Figure \ref{fig:Methods}b).
In a previous publication \cite{Vincent2016}, we showed that this method provided an accurate way of measuring permeability of nanoporous samples. Here, we apply it to the composite, synthetic xylem structures.
In order to perform transpiration experiments, we used samples that were completely filled with water except for a remaining vapor bubble in the serpentine channel.
This situation occurred naturally at the end of a high pressure experiment (see above).
We then placed the samples in a humidity-controlled environment with sub-saturated water vapor (water vapor pressure, $p < p_\mathrm{sat}(T)$).
Due to the local equilibrium at the open edge between liquid water in the nanopores and the external water vapor (see section \ref{sec:Theory}, \emph{Theory}), a reduced (typically negative) pressure developed at the open edge, resulting, after a short transient, in a steady-state flow through the structure \cite{Vincent2016}.
This flow progressively emptied the serpentine channel;
we tracked the resulting motion of the liquid/vapor meniscus in the serpentine channel, using a camera (Point Grey Grasshopper Monochrome camera) a macro lens (AF Micro-Nikkor 60 mm f/2.8), and white, diffuse illumination (Schott ACE 150W light source).
From the velocity of the meniscus, we calculated the mass flow rate through the structure
using Equation (\ref{eq:MassFlowRate}).
For every sample we recorded transpiration-induced flow at a variety of imposed vapor pressures, $p$, in a range high enough to avoid triggering either dewetting (desorption) from the nanopores or cavitation within the microchannels (see section \ref{sec:ResultsDiscussion}, \emph{Results and Discussion}).

\paragraph{Spontaneous imbibition}

We used spontaneous imbibition as a third method to drive transport within our structures.
As we have shown recently \cite{Vincent2017a}, liquid imbibition in a nanoporous medium can occur spontaneously by condensation of the vapor into the nanopores, provided that the vapor pressure of the external vapor is above the threshold for capillary condensation, $p_\mathrm{cond}$.
With our poSi layer, $p_\mathrm{cond} \simeq 0.6 \, p_\mathrm{sat}$ \cite{Vincent2017a}, so that we worked with an imposed vapor pressure above this value, $p = 0.93 \, p_\mathrm{sat}$ to induce condensation and imbibition.
We followed the imbibition dynamics optically, using time-lapse videos obtained with the same imaging system as for transpiration experiments (camera and macro lens, see above).
Because the optical reflectance of the poSi layers changes as a function of water content (see section \ref{sec:ResultsDiscussion}, \emph{Results and Discussion}), we tracked the imbibition front by monitoring relative changes in the gray level of the image, $\Delta I / I$ averaged in the $x$ direction; we then extracted the position of the imbibition front by defining $Y(t)$ as the position at which the signal is at its midpoint between the dry gray level value and the wet value (see \cite{Vincent2017a} for details).
From the displacement, $Y(t)$ of the invading liquid front within the structure, we measured the dynamics of imbibition in our structured samples and estimated the effective permeability of the sample using a modified Lucas-Washburn equation (see section \ref{sec:Theory}, \emph{Theory}).

We performed two types of imbibition experiments. For the first type, we first dried the samples for at least 48 hours in dry air, with a relative humidity (RH) of $6$\%. Then, we placed the samples in a closed box with a humidity of $93$\% RH, set by an unsaturated solution of sodium chloride (NaCl).
For the second type, the samples were evacuated for at least 24 hours in a vacuum chamber. Then, we released water vapor in the chamber to impose a fixed vapor pressure of $p = 0.93 \, p_\mathrm{sat}$ ($93$\% RH). As a result, imbibition occurred with air-filled nanopores and microchannels in the first type of experiments and with evacuated nanopores and microchannels in the second type.

Note that during all imbibition experiments, the microchannels never filled with liquid water,
because the humidity used to trigger condensation in the nanopores was not sufficiently high to induce condensation in the microchannels, nor capillary suction from the nanopores into the microchannel (see Theory, section \ref{sec:CapCondSpontFill})

\section{Theory \label{sec:Theory}}

\subsection{Hypotheses}

Since dimensions of the structures in the $z$-direction ($H_\mathrm{p}$ for the nanoporous layer, $H_\mathrm{v}$ for the microchannels) are much smaller than in other directions, we assume that all gradients leading to transport in the system are in the $x$ and $y$ directions (Figure \ref{fig:Samples}a), and also that there is fast exchange between the microchannels and nanoporous layer in the $z$ direction (local equilibrium).

\subsection{Basic equations}

\paragraph{Transport in the nanopores.}

In all situations considered here (Figure \ref{fig:Methods}), transport within the nanporous layer (poSi) occurred in the liquid state and was thus driven by a liquid pressure gradient $\nabla P \, \mathrm{[Pa/m]}$. Following Darcy's law, the associated mass flux density, $q \, \mathrm{[kg/(m^2.s)]}$ is related to the Darcy permeability, $\kappa_0 \, \mathrm{[m^2/(Pa.s)]}$ through
\begin{equation}
	q_\mathrm{po} = - \rho \kappa_0 \nabla P
	\label{eq:Darcy}
\end{equation}
where $\rho \simeq 10^3 \, \mathrm{kg/m^3}$ is the density of liquid water \cite{Wagner2002}. From our previous studies with poSi layers prepared in the same way, we have, $\kappa_0 \simeq 1.8 \times 10^{-17} \, \mathrm{m^2/(Pa.s)}$ \cite{Vincent2016, Vincent2017a}.

\paragraph{Transport in the microchannels.}


\begin{figure}[ht]
	\includegraphics[scale=0.9]{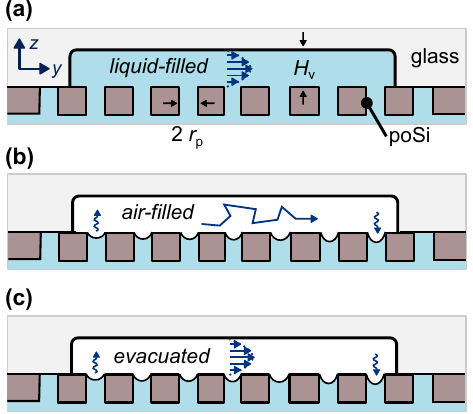}
		\caption{
			\small
			Sketches of transport mechanisms along the length ($y$) of the microchannels in three situations:
			(a) channel filled with liquid (Poiseuille flow of liquid water),
			(b) channel filled with air (diffusive transport of water vapor), and
			(c) evacuated channel (Poiseuille flow of water vapor).
			The sketches depict a cross-section along the length of a microchannel.
			The poSi layer is represented as interconnected pores, drawn with exagerated diameter compared to their actual size.
			The squiggle arrows in (b-c) represent evaporation and condensation at the left and right of the channel, respectively.
			The situation in (a) corresponds to high-pressure or transpiration experiments (Figure \ref{fig:Methods}a-b).
			The situation in (b) and (c) correspond to spontaneous imbibition experiments (Figure \ref{fig:Methods}c), in the air-filled and evacuated cases respectively.
			In all cases, transport in the microchannel occurs in parallel with a Darcy flow in the poSi layer, driven by a pressure gradient (hence, the increasing meniscus curvature from left to right in panels b-c), see Equation (\ref{eq:Darcy}).
			Situations described in panels (a), (b) and (c) correspond to Equations (\ref{eq:q_liq}), (\ref{eq:q_air}) and (\ref{eq:q_vac}), respectively.
		}
	\label{fig:Mechanisms}
\end{figure}


In regions of the sample where microchannels are in contact with the porous layer, transport in the microchannels can occur in parallel to Darcy flow in the nanopores (Figure \ref{fig:Mechanisms}).
This process happens either by division of the liquid flow when the microchannels are filled with liquid (Figure \ref{fig:Mechanisms}a), or by evaporation, transport through the vapor phase, and condensation when the channels are air-filled or evacuated (Figure \ref{fig:Mechanisms}a).

In the cases in Figure \ref{fig:Methods}a-b (high pressure and transpiration), the microstructure was filled with liquid and transport occurred through pressure-driven, Poiseuille-like flow
(Figure \ref{fig:Mechanisms}a).
Since the depth, $H_\mathrm{v}$ of the microchannels was small compared to their other dimensions, we evaluate transport dynamics using the expression of planar Poiseuille flow for the mass flux density (mass flow per unit cross-sectional area, $q_\ell \, \mathrm{[kg.m^{-2}.s^{-1}]}$):
\begin{equation}
	q_\ell = - \rho \frac{H_\mathrm{v}^2}{12 \eta} \nabla P
	\label{eq:q_liq}
\end{equation}
where $\eta \simeq 10^{-3} \, \mathrm{Pa.s}$ is the viscosity of liquid water \cite{Huber2009}.

In imbibition experiments (Figure \ref{fig:Methods}c), however, transport in the microchannels occurred in the vapor phase, either by diffusion through air (air-filled state, Figure \ref{fig:Mechanisms}b)
or as a pressure-driven (convective) flow of pure water vapor (evacuated state, Figure \ref{fig:Mechanisms}c).
From Fick's law, transport by diffusion leads to the following prediction of the mass flux density:
\begin{equation}
	q_\mathrm{v,air} = - \frac{M D}{R T} \nabla p
	\label{eq:q_air}
\end{equation}
where $RT \, \mathrm{[J/mol]}$ is thermal energy, $p \, \mathrm{[Pa]}$ is the partial pressure of water vapor in air, $M = 1.802\times10^{-2} \, \mathrm{kg/mol}$ the molar mass of water and $D \simeq 2 \times 10^{-5} \, \mathrm{m^2/s}$ is the diffusivity of water vapor in air \cite{Massman1998}. In the second situation with an evacuated sample, we expect a Poiseuille flow driven by gradients of vapor pressure:
\begin{equation}
	q_\mathrm{v,vac} = - \rho^\ast \frac{H_\mathrm{v}^2}{12 \eta^\ast} \nabla p
	\label{eq:q_vac}
\end{equation}
where $\rho^\ast = p M / (R T)$ is the density of the vapor and $\eta^\ast \simeq 10^{-5}\, \mathrm{Pa.s}$ its viscosity \cite{Huber2009}.

\paragraph{Liquid-vapor equilibrium (Kelvin equation).}

Lastly, because we consider situations where transport is multiphase (e.g., vapor flow within the microchannels in parallel with liquid flow within the nanopores during imbibition), we need to relate how the liquid and vapor pressure locally relate to one another.
The equality of chemical potentials of the liquid phase and the vapor phase imposes a relation between liquid pressure, $P$ and vapor pressure, $p$, known as Kelvin equation:
\begin{equation}
	P = P_\mathrm{ref} + \frac{RT}{v_\mathrm{m}} \ln \left( \frac{p}{p_\mathrm{sat}} \right) = P_\mathrm{ref} + \Psi(p)
	\label{eq:Kelvin}
\end{equation}
where $v_\mathrm{m} = M / \rho \, \mathrm{[m^3/mol]}$ is the molar volume in the liquid state, $p \, \mathrm{[Pa]}$ the partial (in air) or total (in vacuum) vapor pressure, $p_\mathrm{sat}(T)$ the saturation vapor pressure (e.g., $2811$ Pa at $23\degree$C), and $\Psi = (RT / v_\mathrm{m}) \ln \left( p / p_\mathrm{sat} \right)$ is the \emph{water potential} \cite{Stroock2014,Bacchin2021};
$P_\mathrm{ref}$ is a reference pressure equal to atmospheric pressure when working in air, and equal to $p_\mathrm{sat}$ when working in vacuum. This distinction is anecdotal in our present situations because $P_\mathrm{ref}$ is typically orders of magnitude lower compared to $P$ and can be neglected; $p_\mathrm{sat}$ also changes slightly depending on the air/vacuum context, but again in a negligible manner for our purposes \cite{Vincent2022}.

For convenience, we also define the following dimensionless quantities:
\begin{equation}
	a = \frac{p}{p_\mathrm{sat}}
\end{equation}
the activity (i.e., relative humidity, $0 \leq a \leq 1$) of water vapor and
\begin{equation}
	\epsilon = \frac{v_\mathrm{m} p_\mathrm{sat}}{R T}
\end{equation}
which corresponds to the vapor-to-liquid density ratio at saturation. Using tabulated values of water density and saturation pressure \cite{Wagner2002}, $\epsilon = 1.3 \times 10^{-5}$ at $15 \degree$C and $\epsilon = 2.1 \times 10^{-5}$ at $23 \degree$C.

\paragraph{Capillary condensation and spontaneous filling. \label{sec:CapCondSpontFill}}

Kelvin equation (\ref{eq:Kelvin}) can also be used to estimate the humidity at which pores or channels fill spontaneously by capillary condensation, or by meniscus aspiration from the porous medium.
In order to fill the hydrophilic microchannel, one must form a meniscus in the channel, resulting in a reduced pressure in the liquid due to capillarity ($P \sim - 2 \sigma / H_\mathrm{v} \simeq - 5$ kPa from Laplace's law, with $\sigma = 0.073$ N/m, the surface tension of the liquid).
Condensation becomes possible when this capillary pressure is compatible with Kelvin's equation (\ref{eq:Kelvin}), in other words the typical relative humidity necessary for capillary condensation in the microchannels is $p / p_\mathrm{sat} = \exp \left( - 2 \sigma v_\mathrm{m} / (R T H_\mathrm{v})\right) \simeq 0.99997$, neglecting $P_\mathrm{ref}$.
In other words, condensation in the microchannels is possible only if the imposed relative humidity is extremely close to $100$\% RH, which was not the case in our experiments.
Similarly, Kelvin's equation implies that the pressure in the liquid in the nanopores during imbibition experiments varies between $\Psi(p) \simeq -10$ MPa at the edge of the sample, down to $\Delta P_\mathrm{c} \simeq -70$ MPa at the imbibition front (see below, section \ref{sec:ExpressionsSituations}).
Since these values are much larger in magnitude than the microchannel capillary pressure ($\simeq -5$ kPa), spontaneous invasion of the microchannels was impossible in our experiments, due to the much stronger suction of the nanoporous layer.

\subsection{Relative magnitudes of transport mechanisms \label{sec:RelativeMagnitudeTransport}}

Here, we compare how transport within the microchannels should compare to transport in the nanoporous layer depending on the situation.

Because the cross-sections available for flow are similar in the poSi layer and in the microchannels ($H_\mathrm{p}$ comparable to $H_\mathrm{v}$), we assume below that the ratio of mass flux densities in the porous layer and in the microchannel is a reasonable measure of relative transport between these two elements.

For high pressure flow (Figure \ref{fig:Methods}a) and transpiration (Figure \ref{fig:Methods}b) for which the microchannels were filled with water (Figure \ref{fig:Mechanisms}a), we predict (Equations (\ref{eq:Darcy}) and (\ref{eq:q_liq})) the ratio of fluxes in the channels ($q_\ell$) and in the nanoporous layer ($q_\mathrm{po}$) to be:
\begin{equation}
	\frac{q_\ell}{q_\mathrm{po}} = \frac{H_\mathrm{v}^2}{12 \eta \kappa_0} = \mathcal{O}(10^{9})
	\label{eq:ratio_liq_po}
\end{equation}
so that microchannels can be considered as infinitely conductive relative to the nanoporous matrix in these cases.

For imbibition with evacuated microchannels (Figure \ref{fig:Methods}c, Figure \ref{fig:Mechanisms}c), we predict the ratio of vapor flux in the channels ($q_\mathrm{v,vac}$) to the Darcy flux in the poSi ($q_\mathrm{po}$) to be:
\begin{equation}
	\frac{q_\mathrm{v,vac}}{q_\mathrm{po}} = \frac{(\epsilon a H_\mathrm{v})^2}{12 \eta^\ast \kappa_0} = \mathcal{O}(10^{2})
	\label{eq:ratio_vac_po}
\end{equation}
so that evacuated microchannels filled with pure water vapor should also act as shortcuts for the transport of water, similarly to when they are filled with liquid water.
In deriving the expression in Equation (\ref{eq:ratio_vac_po}), we differentiated Equation (\ref{eq:Kelvin}) to relate vapor pressure gradients in the vapor phase to gradients in liquid pressure in the porous medium ($\nabla p = \epsilon a \nabla P$); we then used this expression for $\nabla p$ in Equation (\ref{eq:q_vac}) to predict $q_\mathrm{v,vac}$.
For the numerical estimate, we have considered $a = p / p_\mathrm{sat}$ to be in the range $0.5$ to $1$, as is typical in our experiments.

Finally, for imbibition with air-filled microchannels (Figure \ref{fig:Methods}c, Figure \ref{fig:Mechanisms}b), we use Equation (\ref{eq:q_air}) and the derivative of Equation (\ref{eq:Kelvin}) to predict the flux in air-filled microchannels ($q_\mathrm{v,air}$) and find the ratio:
\begin{equation}
	\frac{q_\mathrm{v,air}}{q_\mathrm{po}} = \frac{\epsilon^2 a D}{p_\mathrm{sat} \kappa_0} = \mathcal{O}(10^{-1})
	\label{eq:ratio_air_po}
\end{equation}
so that transport in the microchannels should be negligible when they are filled with air such that the flux should be dominated by Darcy flow through the poSi in this scenario.
This situation is similar to that reported recently of negligible transport of water vapor in an air gap above a drying colloidal suspension \cite{Pingulkar2024}.

Since transport in the microchannels is either very large (Equations \ref{eq:ratio_liq_po}-\ref{eq:ratio_vac_po}) or very small (Equation \ref{eq:ratio_air_po}) compared to that in the nanoporous layer, a detailed description of the local exchange and distribution of flow between microchannels and the nanopores situated below in the z-direction (Figure \ref{fig:Mechanisms}) is not necessary:
in the following, we will simply assume that microchannels either completely bypass the nanoporous layer below it when filled with liquid or water vapor, and are invisible (all flow in the nanoporous layer below) when filled with air.

Our design thus presents an interesting situation in which transport can be tuned by changing the contents of the microstructure.
For example, by introducing air in a previously evacuated sample, the microchannels change from being shortcuts for the flow to having virtually no effect on transport.
This change of behavior occurs because
\begin{equation}
	\frac{q_\mathrm{v,air}}{q_\mathrm{po}} \ll 1 \ll \frac{q_\mathrm{v,vac}}{q_\mathrm{po}}.
	\label{eq:ratios_inequalities}
\end{equation}
Later, we demonstrate this switching effect in the context of spontaneous imbibition.

Note that varying the ratio $H_\mathrm{v} / H_\mathrm{p}$ is in principle another way to tune the relative flow between microchannels and porous layer.
For example, from Equation (\ref{eq:ratio_vac_po}), the mass flux density decreases rapidly in the evacuated microchannel when its height is decreased, thus by lowering this value enough, one should be able to make it similar to the mass flux density in the porous layer; this choice would result in a similar behavior between evacuated and air-filled samples (negligible flow in the microchannels).
We have not explored this possibility in the present study.
This observation shows the importance of the choice of physical and geometrical parameters for both nanoporous layer and microchannel for the switching and tunable behavior to be possible.

\subsection{Effective permeability of the composite structures\label{sec:EffectivePermComposite}}

We now estimate the impact of the microstructure on transport depending on its geometry. Following the estimates from the previous section, we consider 1) that all samples are equivalent to a pure nanoporous medium without any microstructure (\emph{Blank}) when the microchannels are filled with air (based on Equation \ref{eq:ratio_air_po}); 2) that each microchannel is infinitely conductive in all other situations (microchannels filled with liquid or pure water vapor), based on Equations (\ref{eq:ratio_liq_po}) and (\ref{eq:ratio_vac_po}).

\paragraph{Definitions.}

From Darcy's law (Equation \ref{eq:Darcy}), the mass flow rate, $Q \, \mathrm{[kg / s]}$ driven by a gradient of pressure, $\Delta P / L$ in the \emph{Blank} sample is $Q_0 = \mathcal{S} q_\mathrm{po}$, or
\begin{equation}
	Q_0 = \rho \kappa_0 \mathcal{S} \frac{\Delta P}{L}
	\label{eq:DarcyBlank}
\end{equation}
where $\mathcal{S} = W \times H_\mathrm{p}$ is the cross-section area of the porous layer.
By analogy, for a sample with a microstructure, we define the effective permeability, $\kappa$ such that the mass flow rate through the sample is
\begin{equation}
	Q = \rho \kappa \mathcal{S} \frac{\Delta P}{L}
	\label{eq:DarcyEff}
\end{equation}
i.e., we consider the effect of the microstructure without considering details of its geometry, e.g., the fact that it is not actually embedded in the nanoporous layer. In other words, we measure the increase of permeability in a sample compared to the reference (\emph{Blank}) by the increase of the observed mass flow rate in similar experimental situations:
\begin{equation}
	\frac{\kappa}{\kappa_0} = \frac{Q}{Q_0}.
	\label{eq:DarcyRatio}
\end{equation}


\begin{figure*}[ht]
	\includegraphics[scale=1]{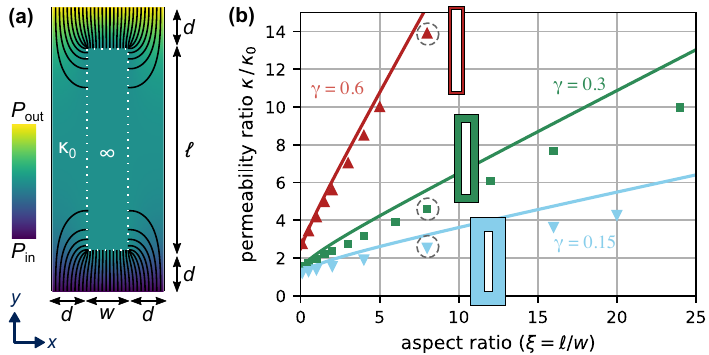}
		\caption{\small
		Numerical and analytical predictions of steady-state flow through the composite matrix, as a function of the microchannel aspect ratio ($\xi =  \ell / w$) and of the surface density of channels ($\gamma$).
		(a) Pressure field (colormap) and streamlines (black lines) for the unit cell of the composite porous medium, obtained from numerical simulations (here with $\gamma = 0.3$, $\xi = 2.2$); inside the white dotted line is the microchannel over the porous layer and outside this line is just the porous substrate.
		In this prediction, we assumed that the microchannels had infinite permeability, a reasonable approximation when the microchannels are filled with liquid water or pure water vapor (see \emph{Theory}, Equations (\ref{eq:ratio_liq_po}) and (\ref{eq:ratio_vac_po})).
		(b) Comparison of numerical and analytical predictions of the effective permeability ratio $\kappa / \kappa_0$ as a function of $\xi$ for different values of $\gamma$.
		Symbols represent predictions of the numerical simulations as in (a), and lines represent approximate, analytical predictions based on a resistors-in-series approach (Equation \ref{eq:AnalyticalPermeabilityGammaXi}).
		Sketches show the appearance of a unit cell for the three different values of $\gamma$, at an aspect ratio, $\xi = 8$;
		the data corresponding to these three sketches is circled in gray (dashes).
		See Appendix \ref{sec:AppendixMaxwell}, Figure \ref{fig:SimulationsMaxwell}  for a similar comparison between simulation results and a Maxwell-Garnett-like approach (Equation \ref{eq:AnalyticalPermeabilityMaxwell}).
		}
	\label{fig:Simulations}
\end{figure*}


\paragraph{Simulations.}

We have evaluated $\kappa / \kappa_0$ for different microstructure geometries with numerical simulations, assuming that areas directly covered by the microstructure (microchannel) were infinitely conductive (Figure \ref{fig:Simulations}a).
We predicted the pressure-driven transport through the multi-scale porous media by solving the time-dependent, poroelastic diffusion equation for pressure by finite difference for a single unit cell of the patterned structure; this unit cell is shown in Figure \ref{fig:Simulations}a.
This approach is an example of effective medium theory \cite{Choy2016}, where the explicit transport is resolved over a large enough (i.e. statistically representative) domain to include the diversity of material conductance present in the full system.
The well-ordered, periodic microstructure of our system allows for the single unit cell of the patterned structure to serve as such a domain \cite{Renard1997}.

In our numerical solution, we divided the unit cell into two domains: the microchannel (inside white dotted line in Figure \ref{fig:Simulations}a) and the nanoporous domain (outside white dotted line). Consistent with its high conductance (Equations (\ref{eq:ratio_liq_po}) and (\ref{eq:ratio_vac_po})), we treated the microchannel as a "lumped capacitance" with a single, uniform pressure at each time step. In the porous domain, we solved the poroelastic diffusion equation using an explicit finite difference scheme.
The boundary conditions imposed on the porous domain were no flux (right and left), fixed pressure $P_\mathrm{in}$ at the bottom, fixed pressure $P_\mathrm{out}$ on top, and uniform pressure along the boundary between the microchannel and the rest of the domain (Figure \ref{fig:Simulations}a, white dotted line).
We have provided a detailed description of this numerical approach elsewhere (see Supplementary Information of \cite{Vincent2014}, and section 4.2.1 of \cite{Huber2017}).
We allowed the solution to reach steady state for a fixed difference in pressure imposed on the
bottom ($P_\mathrm{in}$) and top
boundaries ($P_\mathrm{out}$) to assess the steady-state mass flow rate, $Q$ and thus the effective permeability increase from Equation (\ref{eq:DarcyRatio}).

We numerically calculated $\kappa / \kappa_0$ for a variety of geometries with different surface coverage ($\gamma$) of the microstructure and with varying aspect ratio of the microchannels (see Figure \ref{fig:Simulations}b, symbols). The $\gamma = 0.3$ case is close to our experimental samples (see Table \ref{tab:Samples}); however our fabrication procedure induced variations in the values of $\gamma$ for the different samples. Consequently, we also ran specific numerical simulations with each actual sample geometry, as measured on optical micrographs; the corresponding values are indicated in the last column in Table \ref{tab:Samples}.

\paragraph{Analytical estimate (effective medium approach).}

We also used effective medium approaches \cite{Choy2016} to derive analytical formulas, in order to estimate the value of the effective permeability of our structures, without having to perform numerical simulations.

First, we adapted an effective medium approach originating from J. C. Maxwell and J. C. M. Garnett \cite{Garnett1904,Choy2016}, and extended by Zimmerman to evaluate the effective permeability of two-dimensional media with elliptical inclusions \cite{Zimmerman1996}.
In our approach, we assume that the effect of the array of microchannels was identical to that of an ensemble of ellipses with the same aspect ratio, $\xi$.
We adapted Zimmerman's calculation to account for aligned ellipses with infinite permeability , and found
\begin{equation}
	\frac{\kappa}{\kappa_0}
	=
	\frac{
		2 + (1 + \xi) \gamma
	}
	{
		2 - (1 + \xi) \gamma
	}
	\label{eq:AnalyticalPermeabilityMaxwell}
\end{equation}
(details of the derivation can be found in Appendix \ref{sec:AppendixMaxwell}, and in \cite{Huber2017}).
Equation (\ref{eq:AnalyticalPermeabilityMaxwell}) gave satisfactory agreement with our simulation results, but only for low areal fractions, $\gamma$, and moderate aspect ratios, $\xi$ (see Appendix \ref{sec:AppendixMaxwell}, Figure \ref{fig:SimulationsMaxwell}).
This observation is not surprising, given that Zimmerman's calculation is a first-order approach which performs best at low densities of inclusions \cite{Zimmerman1996}.
As a result, we also used another approach better suited for large areal coverages and large aspect ratios.

In this approach, we account only for resistance to flow along the y-axis in the sections of length, $d$ without a microchannel (see Figure \ref{fig:Simulations}a);
we assume that the microchannel short-circuits the flow through the section that contains a microchannel, of length, $\ell$.
This treatment leads to the following prediction for the effective conductance:
\begin{equation}
	\frac{\kappa}{\kappa_0} = \frac{\ell + 2d}{2d} = \frac{1}{\delta}
	\label{eq:AnalyticalPermeability}
\end{equation}
where $\delta = 2d / (\ell + 2d)$ is the ratio between the distance traveled in poSi only and the total axial path length across a unit cell. We have used Equation (\ref{eq:AnalyticalPermeability}) it to generate the analytical estimates in Table \ref{tab:Samples}.

If the inputs are the aspect ratio of the microchannels ($\xi$) and the surface coverage ratio ($\gamma$) instead of $\ell $ and $d$, some more algebra is needed to relate these quantities. With our design keeping the same half-separation $d$ in both directions around the microchannels (Figure \ref{fig:Samples}, Figure \ref{fig:Simulations}a), one can relate $\delta$ in a unique manner to $\xi$ and $\gamma$ through $\delta = f(\xi, \gamma) - \sqrt{f(\xi, \gamma)^2 + \gamma - 1}$, with $f(\xi, \gamma) = 1 + \frac{\gamma}{2} (\xi - 1)$. As a result,from Equation (\ref{eq:AnalyticalPermeability}),
\begin{equation}
	\frac{\kappa}{\kappa_0}
	=
	\frac{
		1
		}
		{
		f(\xi, \gamma) \left[
			1 - \sqrt{1 - \frac{1 -\gamma}{f(\xi, \gamma)^2}}
			\right]
		}
	\label{eq:AnalyticalPermeabilityGammaXi}
\end{equation}
We used Equation (\ref{eq:AnalyticalPermeabilityGammaXi}) to generate the lines in Figure \ref{fig:Simulations}b.

Comparison with simulations (symbols in Figure \ref{fig:Simulations}b) indicates that our analytical approach (Equations \ref{eq:AnalyticalPermeability}-\ref{eq:AnalyticalPermeabilityGammaXi}) performs well at high surface coverage of the microstructure (e.g. $\gamma=0.6$, red triangles and curve), with more significant over estimations of $\kappa / \kappa_0$ at lower densities of microchannels.
As an example, if we evaluate Equation (\ref{eq:AnalyticalPermeability}) (equivalently, Equation \ref{eq:AnalyticalPermeabilityGammaXi}) for the specific geometries that we have used experimentally in this study, we find that it overpredicts the permeability ratio by $15\--20 \%$ (see Table \ref{tab:Samples}).
Graphical comparison between Equations (\ref{eq:AnalyticalPermeability}-\ref{eq:AnalyticalPermeabilityGammaXi}) and numerical simulations is also available in the \emph{Global results} discussion a the end of the article (see Figure \ref{fig:GlobalResults}a).

\subsection{Expressions corresponding to the different experimental situations \label{sec:ExpressionsSituations}}

\paragraph{High pressure flow.}

High pressure experiments (Figure \ref{fig:Methods}a) correspond to a steady-state situation with an external pressure, $P$, pushing the liquid towards the serpentine channel (at pressure, $P_\mathrm{s}$).
The typical magnitude of $P$ was $\sim 10$ MPa ($10^7$ Pa).
Since the sample was evacuated before high pressure experiments, the gas pressure in the bubble at the end of the serpentine channel was the saturation pressure of water: $P_\mathrm{s} = p_\mathrm{sat}(T) \simeq 2$ kPa $\sim 10^3$ Pa (we can neglect here the effect of curvature on the vapor pressure since the meniscus dimensions are in the micrometer scale).
Also, the Laplace pressure due to the curvature of the liquid-vapor interface between bubble and liquid in the channel is on the order of $- 2 \sigma / H_\mathrm{v} \simeq - 5$ kPa.
Thus, $P_\mathrm{s} \sim 10^3 \, \mathrm{Pa} \ll P \sim 10^7 \, \mathrm{Pa}$ so that the pressure difference between the edge of the sample and the serpentine channel is, to very good approximation, $\Delta P = - P$. From Darcy's law (Equation \ref{eq:DarcyEff}), it follows that the mass flux is
\begin{equation}
	Q = - \rho \kappa \mathcal{S} \frac{P}{L}
	\label{eq:Filling}
\end{equation}
Note that the mass flow rate is negative, corresponding to an inwards flux (towards $y<0$).

\paragraph{Transpiration.}

Similarly to high pressure experiments, transpiration (Figure \ref{fig:Methods}b) is a steady-state situation but with a negative pressure driving the flow towards the open edge of the sample instead of an external overpressure driving the flow towards the backend of the structure. This negative pressure originates from the local equilibrium between the liquid and vapor phases at the open edge and is mediated by the curvature of the exposed nanoscale menisci \cite{Vincent2014}. Due to this equilibrium, the liquid pressure at the open edge was imposed by the external water vapor pressure through Kelvin equation (Equation \ref{eq:Kelvin}). The magnitude of the Kelvin pressure was $\sim 10$ MPa in our experiments, so that the reference pressure, $P_\mathrm{ref}$ and the serpentine pressure $P_\mathrm{s}$ can again be neglected. As a result:
\begin{equation}
	Q = - \rho \kappa \mathcal{S} \frac{\Psi (p)}{L}
	\label{eq:Transpiration}
\end{equation}
where
\begin{equation}
	\Psi (p) = \frac{RT}{v_\mathrm{m}} \ln \left(\frac{p}{p_\mathrm{sat}}\right)
	\label{eq:WaterPotential}
\end{equation}
is the (negative) Kelvin pressure, which also corresponds to the \emph{water potential} at the open edge of the sample (see Equation \ref{eq:Kelvin}).
The resulting flux is positive, i.e., with a $y>0$ direction (towards the open edge, see Figure \ref{fig:Methods}c).

\paragraph{Imbibition by capillary condensation.}

In the situation of Figure \ref{fig:Methods}c, water vapor condenses at the open edge and the condensed liquid in the nanopores invades the structure, driven by the capillary pressure, $\Delta P_\mathrm{c} = - 2 \sigma \cos \theta / r_\mathrm{p}$, where $r_\mathrm{p}$ is the nanopores radius and $\theta$ the contact angle of water on the pore walls.
From our previous studies \cite{Vincent2016,Vincent2017a}, $\Delta P_\mathrm{c} \simeq -70$ MPa for the nanoporous silicon layer.
This situation is similar to regular spontaneous imbibition, i.e., capillary invasion when the sample edge is soaked in bulk liquid water; this process is described by the Lucas-Washburn equation \cite{Lucas1918,Washburn1921a}.
However, in the present situation where imbibition is triggered by capillary condensation, the liquid-vapor equilibrium at the open edge introduces another capillary pressure, $\Psi (p)$ (through the Kelvin equation, see Equation \ref{eq:Kelvin}) that resists the invasion \cite{Vincent2017a}. As a result the invasion dynamics (front position, $Y (t)$) is described by a modified Lucas-Washburn equation that takes into account the competition of these two capillary pressures, as we have shown previously \cite{Vincent2017a}:
\begin{equation}
	Y (t) = \sqrt{\frac{2 \kappa}{\phi} \left( \Psi(p) - \Delta P_\mathrm{c} \right) t}
	\label{eq:Imbibition}
\end{equation}
where $\phi$ is the porosity of the nanoporous layer; $\omega = 2 \kappa (\Psi - \Delta P_\mathrm{c}) / \phi = Y^2 / t$ is the Lucas-Washburn "velocity" of imbibition (homogeneous to a diffusivity $\mathrm{[m^2 / s]}$) \cite{Vincent2017a}, and depends on the humidity imposed around the sample through $\Psi (p)$.
In our experiments, we used a fixed value of relative humidity, $p / p_\mathrm{sat} = 0.93$ (see \emph{Methods}, section \ref{sec:MethodsExperiments}), resulting in $\Psi(p) \simeq -9.9$ MPa from Equation (\ref{eq:Kelvin}).
As a result, the imbibition flow was driven by a gradient of pressure going from $\Psi(p) \simeq -9.9$ MPa at the sample edge, down to $\Delta P_\mathrm{c} \simeq -70$ MPa at the imbibition front.

\section{Results and Discussion \label{sec:ResultsDiscussion}}

\subsection{High pressure flow \label{sec:ResultsPBomb}}

We used high pressure, steady-state flow in our various composite structures to estimate the permeability associated with samples containing microstructure (samples \emph{Square}, \emph{Short}, \emph{Long}, \emph{Vein}), relative to that of a purely nanoporous sample (\emph{Blank}, permeability $\kappa_0$).
From the observed filling velocity  $U = dX / dt$ of the meniscus in the serpentine channel (see \emph{Methods} and Appendix \ref{sec:AppendixPBomb}, Figure \ref{fig:ForcedFilling}), and Equations (\ref{eq:MassFlowRate}) and (\ref{eq:Filling}) we estimated the permeability ratio $\kappa / \kappa_0$, where the index $0$ refers to the reference sample (\emph{Blank}).
We report and discuss the extracted values of the permeability ratios in the general discussion (see below, section \ref{sec:GlobalResults} and Figure \ref{fig:GlobalResults}b).

\subsection{Transpiration}


\begin{figure*}[ht]
	\includegraphics[scale=1]{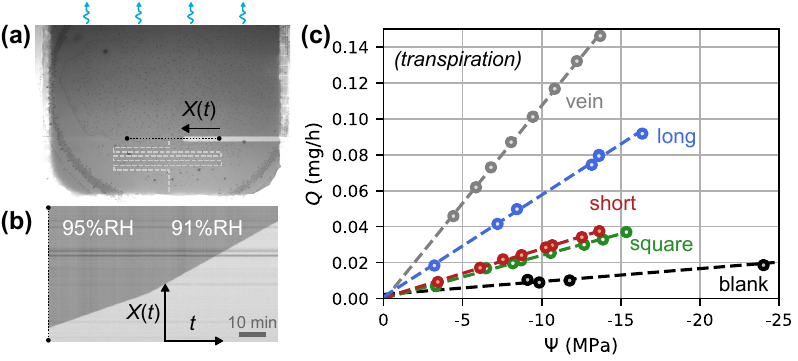}
		\caption{\small
		Transpiration-driven flow (see Figure \ref{fig:Methods}b).
		(a) Picture of the \emph{Long} sample during a transpiration experiment (contrast has been enhanced for clarity). In this particular case, the liquid-vapor meniscus was located in the distribution channel (i.e. not in the serpentine part -- sketched with white dashes); however we could still track the expanding bubble to measure transpiration-driven flow.
		Note that no cavitation can be seen, and this was the case for all samples down to $\Psi \simeq -15$ MPa.
		(b) Reslice (stack of the pixels along the black line in panel (a) for all images in the recorded sequence), showing the expansion of the bubble in the channel over time as the humidity is switched from $a=0.95$ to $a=0.91$; the slope of the visible line separating two areas of different gray levels is proportional to the velocity of the meniscus, and hence to the flow rate in the sample.
		(c) Extracted mass flow rate for all samples, as a function of the negative pressure (water potential, $\Psi$, deduced from the imposed humidity, Equation \ref{eq:WaterPotential}), imposed by the liquid-vapor equilibrium with a subsaturated vapor at the open edge of the sample.
		The slope of the lines is proportional to the ratio, $\kappa / L$ between the effective permeability and the length of the composite structure (Equation \ref{eq:Transpiration}).
		Note that the \emph{Vein} sample was 1.5 times longer than the other samples.
		The measurements for the \emph{Blank} sample extends to $\Psi = -70$ MPa (data not shown but included in the fitting procedure).
		}
	\label{fig:Transpiration}
\end{figure*}


Figure \ref{fig:Transpiration} presents our measurements of transpiration flow rates for each sample geometry as a function of water potential (Equation \ref{eq:WaterPotential}) of the vapor to which the sample was exposed.
After each step in vapor pressure, we observed clear steady-state flows (as measured by the emptying speed, $U = \mathrm{d}X / \mathrm{d}t$ of the serpentine channel) through the structure, after short transients of a few minutes.
Lowering $p$ below $p_\mathrm{sat}$ increased the magnitude of the water potential $\Psi (p)$, controlling the negative pressure driving the flow to the open edge.
As seen in Figure \ref{fig:Transpiration}c, we observed for each sample a linear relationship between the flow rate and $\Psi (p)$ (i.e. a logarithmic dependence of $Q$ on $p$), as predicted by Equation (\ref{eq:Transpiration})-(\ref{eq:WaterPotential}).
We extracted permeability ratios, $\kappa / \kappa_0$ (see section \ref{sec:EffectivePermComposite}) using linear fittings of the experimental data and Equations (\ref{eq:MassFlowRate}) and (\ref{eq:Transpiration}).
See below for a general discussion of these results (section \ref{sec:GlobalResults} and Figure \ref{fig:GlobalResults}b).

We note here that transpiration experiments imply a pressure in the fluid that is negative: $P \simeq \Psi(p) < 0$ near the sample edge, see Equations (\ref{eq:Kelvin}) and (\ref{eq:WaterPotential}).
Bulk fluid at negative pressure is under mechanical tension and is metastable with respect to the spontaneous nucleation of bubbles (\emph{cavitation}) \cite{Stroock2014,Vincent2022}.
In the experiments reported here, we could reach pressures down to $\Psi \simeq -15$ MPa ($p / p_\mathrm{sat} \simeq 0.9$) without cavitation in the microchannels.
These values are consistent with studies with similar devices, showing that the cavitation threshold of water in porous silicon/glass assemblies is the range of $-15$ to $-30$ MPa \cite{Vincent2014,Pagay2014,Chen2016a,Vincent2019}; in fact this range is typical of water cavitation in a variety of systems \cite{Caupin2013,Vincent2022}.
Plants themselves routinely operate with water at negative pressures of a few MPa in magnitude \cite{Cochard2006,Stroock2014}.

For the \emph{Blank} sample with no macroscopic structure, we could extend the measurements down to $\Psi \simeq -70$ MPa ($p / p_\mathrm{sat} \simeq 0.6$), because this sample only contained nano-confined fluid, which can be stable even at negative pressure \cite{Caupin2008}. We expect the liquid phase in the nanoporous layer to start being unstable (inducing dewetting/desorption) for $\Psi(p) < -70$ MPa, \cite{Vincent2016,Vincent2017a}.

\subsection{Imbibition \label{sec:ImbibitionResults}}


\begin{figure*}[ht]
	\includegraphics[scale=1]{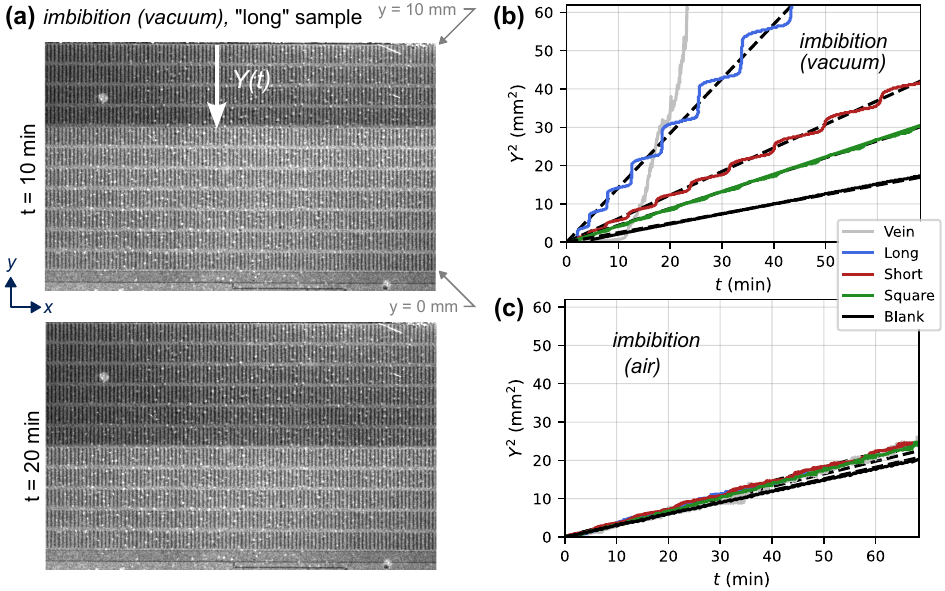}
		\caption{\small
		Dynamics of spontaneous imbibition by capillary condensation (see Figure \ref{fig:Methods}c).
		(a) Images recorded during the experiments in the \emph{Long} sample when evacuated. The darker zone on top of the images is the area wetted by the fluid (contrast has been enhanced for clarity).
		See Supplemental Material for the corresponding complete video (Appendix \ref{sec:AppendixVideo}).
		(b-c) Dynamics of imbibition for all samples when evacuated (b) and air-filled (c), represented by the squared position, $Y^2$, of the wetting front as a function of time (see Figure \ref{fig:Methods}c).
		Different colors represent the different sample geometries, and black lines represent linear fits, indicating Lucas-Washburn-like dynamics on average (Equation \ref{eq:Imbibition}).
		}
	\label{fig:Imbibition}
\end{figure*}


Figure \ref{fig:Imbibition}a (see also Supplemental Material for the corresponding movie) shows images obtained during an imbibition experiment, in a case in which the sample was initially evacuated of air.
The wetting front is clearly visible as a darker zone on the images, allowing us to analyze the imbibition dynamics and extract the front position as a function of time, $Y(t)$, using image processing (see \emph{Methods}).

Figure \ref{fig:Imbibition}b presents the measured dynamics for all samples when they are evacuated.
Clearly, there is a strong enhancement of the imbibition speed when the aspect ratio of the microchannels is increased. This effect occurs because the evacuated channels behave as shortcuts for the transport of water in the structure through the efficient convective flow of pure water vapor, in parallel with the less efficient liquid flow in the nanoporous layer (see \emph{Theory}, Equation \ref{eq:ratio_vac_po}).

The data also displays an apparently intermittent dynamics (clearly visible e.g. for the \emph{Long} sample -- blue curve in Figure \ref{fig:Imbibition}b), which contrasts with the usual continuous progression predicted by Lucas-Washburn theory ($Y \sim \sqrt{t}$) for the wetting front into a homogeneous porous medium. The visible "jumps" in the dynamics (see \emph{Supplementary Movie}) occur when the front advances quickly across a row of microchannels. However, over longer times, the movement of the front follows a trend with $Y^2 \propto t$ scaling, as shown by the dashed, black lines in Figure \ref{fig:Imbibition}b;
these lines were obtained by a linear, least-squares fit of the data.

The Lucas-Washburn like dynamics that we observe when considering the dynamics on dimensions much larger than the length of a single row of microchannels can be explained in the standard way that applies to a imbibition in a uniform porous medium \cite{Lucas1918,Washburn1921a,Vincent2017a}.
First, the driving force for the imbibition flow as described by Equation (\ref{eq:Imbibition}) is the capillary pressure difference, $\Psi(p) - \Delta P_\mathrm{c}$ in the nanoporous layer, which is not impacted by the presence of the microstructure.
Second, the response to this driving force is determined by the permeability of the wet zone between the open edge and the imbibition front; as the front progresses and more and more rows of microchannels are wetted, the behavior of this wet zone approaches the behavior of a homogeneous medium, with a permeability that should match that measured with other methods (high pressure and transpiration) and that predicted by our numerical simulations (see Figure \ref{fig:Simulations} and Table \ref{tab:Samples}).
The hydraulic resistance of this wetted zone grows in proportion to its length, $Y$ such that the speed decreases $\sim 1 / Y$ and the front advances $\sim \sqrt{t}$.

From the slope, $\omega$ of the linear fits in Figure \ref{fig:Imbibition}b, we have extracted effective permeability ratios, $\kappa / \kappa_0 = \omega / \omega_0$ using Equation (\ref{eq:Imbibition}), assuming that the balance of capillary pressures driving the flow ($\Psi(p) - \Delta P_\mathrm{c}$) was constant across experiments; this assumption originates from the use of the same condensation humidity in all imbibition experiments ($\Psi(p)$ constant) and of the pore capillary pressure being dictated by the poSi layer independently of the microstructure superimposed to it ($\Delta P_\mathrm{c}$ constant).
We compare the obtained values to that of other methods and simulations in the \emph{Global results} (section \ref{sec:GlobalResults} and Figure \ref{fig:GlobalResults}b) below.

Noticeably, the \emph{Vein} sample displayed imbibition dynamics (Figure \ref{fig:Imbibition}b) that could not be easily fitted with a Lucas-Washburn equation; this is consistent with the previous remarks, given the fact that the \emph{Vein} sample only contained one row of microchannels that spanned nearly the entire length of the sample;
as such, its dynamics cannot be accounted for with an averaged, effective medium approach as the detailed dynamics of filling of a unit cell accounts for most of the observed dynamics.
Notably, the \emph{Vein} sample started with much slower dynamics than all other samples, even \emph{Blank}.
We hypothesize that this slow initial propagation occurred because the very long microchannel acted as a sink for water molecules, and the observed delay corresponded to the time required to saturate the porous area below the microchannel.
In Appendix \ref{sec:AppendixFillingTime}, we show that we expect the timescale of such a process to be $\tau \sim 2 d L_\mathrm{cell} / \omega_0$, with $L_\mathrm{cell} = \ell + 2d$ the length of the unit cell, and $\omega_0$ the Lucas-Washburn coefficient of the nanoporous layer (see \emph{Theory}, Equation \ref{eq:Imbibition}). For the porous silicon that we have used in our samples, and with an applied humidity of $p / p_\mathrm{sat} = 0.93$, we expect $\omega_0 \simeq 5 \times 10^{-9} \, \mathrm{m^2 / s}$ \cite{Vincent2017a}; with the dimensions of the \emph{Vein} sample (see Table \ref{tab:Samples}); with these values, we estimate $\tau \sim 25$ min, consistent with the dynamics observed in Figure \ref{fig:Imbibition}b.
In comparison, similar estimates with the other samples with microstructures (\emph{Square}, \emph{Short}, \emph{Long}) yield $\tau < 20$ s, much shorter than the observed imbibition dynamics in Figure \ref{fig:Imbibition}b. As a result, we do not expect the filling time of individual rows of cells to be limiting for these other geometries.

Interestingly, when doing experiments with the exact same structures but containing air instead of pure water vapor, the effects related to the microstructure (speeding up of the invasion front, intermittent dynamics) were completely lost and all samples displayed the same continuous Lucas-Washburn dynamics as for a purely nanoporous sample (Figure \ref{fig:Imbibition}c).
This observation is consistent with our prediction (see \emph{Theory}, Equations \ref{eq:ratio_vac_po}-\ref{eq:ratio_air_po}) that the vapor transport mechanism in the microstructure changes from convective (evacuated) to diffusive (in air) and becomes less efficient than liquid transport in the nanoporous layer.
Thus, the microstructure should have no visible effect on imbibition dynamics when filled with air at ambient pressure.

As a result, a single sample can exhibit vastly different transport properties depending on how it was prepared (evacuated vs. air-filled). In other words, samples that behave the same way with respect to imbibition when they are filled with air can "reveal" their microstructure when evacuated, and a single geometry can display a wide range of water transport dynamics.

\subsection{Global results\label{sec:GlobalResults}}


\begin{figure*}[ht]
	\includegraphics[scale=1]{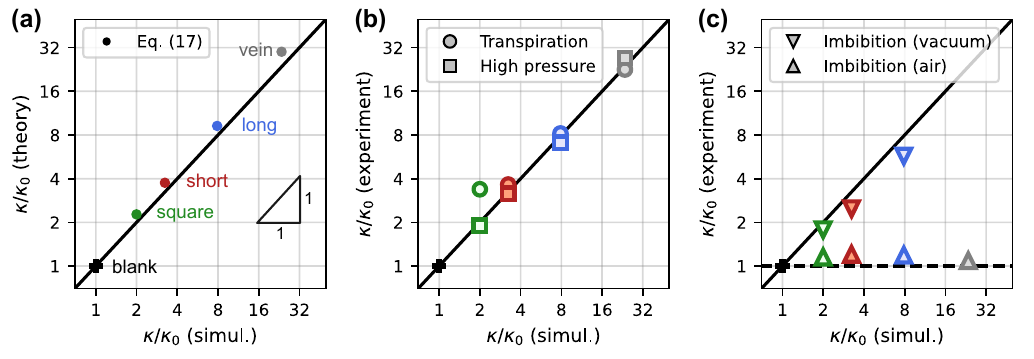}
		\caption{\small
		Predicted and measured effective permeabilities for the different sample designs.
		All graphs have as the x-axis the permeability ratio, $\kappa / \kappa_0 \mathrm{(simul.)}$, which is predicted by numerical simulations, assuming that the microchannels are infinitely conducting (see section \ref{sec:Theory}, \emph{Simulations} paragraph).
		We use the value of $\kappa / \kappa_0 \mathrm{(simul.)}$ as a reference measure for the different samples, using the values listed in Table \ref{tab:Samples}.
		We compare this reference to values obtained with
		(a) the approximate analytical approach (Equation \ref{eq:AnalyticalPermeability}, Table \ref{tab:Samples}),
		(b) steady-state flow measurements (transpiration, circles; high pressure, squares).
		(c) spontaneous imbibition in evacuated samples (downward triangles) and in air-filled samples (upward triangles);
		no data is shown for evacuated imbibition of the vein sample, because the dynamics in this situation is not Lucas-Washburn-like.
		In all panels, the continuous black line corresponds to a linear 1:1 correlation.
		The flat, dashed line in (c) corresponds to $\kappa / \kappa_0 = 1$ (effective permeability independent of microstructure geometry), as predicted when microchannels contribute negligibly to transport;
		this case is expected when microchannels are filled with air (see Equations \ref{eq:ratio_air_po}-\ref{eq:ratios_inequalities}).
		All panels use a $\log_2$ scale for both x and y axes.
		Uncertainties are smaller than symbol size.
		}
	\label{fig:GlobalResults}
\end{figure*}


Figure \ref{fig:GlobalResults} summarizes the effective permeability measurements obtained from numerical simulations, analytical estimates, and from all the experimental approaches developed above.
Figure \ref{fig:GlobalResults}a compares the approximate, analytical estimate (Equation \ref{eq:AnalyticalPermeability}) to numerical simulations results, and is a graphical representation of the rightmost two columns in Table \ref{tab:Samples}.
While we use finite difference simulations as a reference, it is sometimes more convenient to use simpler approaches for design rules-of-thumb. As can be seen in Figure \ref{fig:GlobalResults}a (and more generally in Figure \ref{fig:Simulations}),
Equations (\ref{eq:AnalyticalPermeability}-\ref{eq:AnalyticalPermeabilityGammaXi})
provide reasonable estimates of what can be expected for a particular geometry, with typically less than $20 \%$ error in $\kappa / \kappa_0$ when the surface coverage of the microchannels exceeds $0.3$.

Figure \ref{fig:GlobalResults}b presents the permeability ratios estimated from steady-state flow with liquid-filled microstructure, using two different methods (high pressure and transpiration), for all samples.
The agreement with the predictions from our simulations is excellent ($<15 \%$ deviations), except for the transpiration result for the \emph{Square} sample (experiment $\sim 70 \%$ larger than expected).
Since this outlier measurement was historically the last one in the series made with the \emph{Square} sample and all previous measurements (high pressure, evacuated imbibition, imbibition in air) were consistent with expectations, there is a possibility that this sample was damaged before the last run, e.g., developed a leaky bond between glass and silicon.

Finally, Figure \ref{fig:GlobalResults}c presents effective permeabilities obtained from experiments of spontaneous imbibition dynamics. In these experiments the microstructure was not liquid-filled anymore, but filled with pure water vapor (evacuated case) or with a mixture of water vapor and air (air-filled case).

For imbibition in evacuated samples (Figure \ref{fig:GlobalResults}c, downwards triangles), the permeability ratio extracted from the experiments is very close to that calculated with our simulations (Figure \ref{fig:Simulations}), which assume infinite permeability in the microchannels.
We note however a trend of slower dynamics than predicted (from $10 \%$ slower for the \emph{Square} sample, up to $27 \%$ slower for the \emph{Long} sample), which might be a sign that the permeability of vapor-filled elements should be considered large but not infinite. It is also possible that the complex, intermittent dynamics observed on the local scale (Figure \ref{fig:Imbibition}b) plays a larger role in the global front progression than assumed in our effective medium approach.

For imbibition in the same samples filled with air (Figure \ref{fig:GlobalResults}c, upwards triangles), as discussed previously (section \ref{sec:ImbibitionResults} and Figure \ref{fig:Imbibition}c), the effect of the microstructure vanishes and all samples become equivalent. This is because, given the permeability of the poSi layer and the geometry of our microstructure, diffusive water vapor transport through air in the microchannels is significantly less efficient than liquid transport in the nanopores (Equation \ref{eq:ratio_air_po}). Note however that all samples display $\kappa / \kappa_0$ slighly above 1, which might suggest that air-filled elements have a small, but non-zero conductivity. This remark is consistent with our order-of-magnitude estimates in the \emph{Theory} section (Equations \ref{eq:ratio_liq_po}-\ref{eq:ratio_vac_po}).

We conclude that our proposition that liquid-filled or vapor-filled microchannels act as local shortcuts for water transport is consistent with our experimental results, and that we can accurately predict the associated increase of effective permeability across a structure containing a large arrays of these elements, as a function of the geometry of the individual components.

We have also shown that the design of our composite structures allow us, for a single geometry, to activate or deactivate transport in the microstructure by changing its filling state (evacuated vs. air). Note that this transition as a function of filling state occurs because of the specific combination of geometries and physical properties of our composite structures. Different choices, e.g., of channel depth, or in the typical pore sizes / porosities of the nanoporous layer (i.e. in its permeability, $\kappa_0$) might not satisfy the inequalities described in the \emph{Theory} section (Equation \ref{eq:ratios_inequalities}), which are necessary to observe this change in behavior.

\section{Conclusion}

In this study, we have investigated water transport in composite structures that combine a nanoporous layer and regular arrays of microchannels, inspired by the vascular structures of plants (xylem).
We have shown that we could accurately predict transport properties as a function of the geometry of the microstructure (aspect ratio of its elements) in a variety of situations: high pressure, transpiration, imbibition in air, imbibition in vacuum.

In all these experiments, the nanoporous layer is always filled with liquid water, while the contents of the microstructure depends on the situation: liquid water during high pressure flow and transpiration, water vapor either pure or mixed with air during imbibition.
We have shown that we can take advantage of the different transport mechanisms associated with each situation to tune the transport properties of the structures.
In particular, a single structure can switch from a highly conductive state to a low-conductance one by introducing air within the elements.
Conversely, evacuating the air or refilling the microchannels with liquid makes the structure highly conductive again. As a result, structures with different geometries can be identical with respect to transport when filled with air but reveal vastly different transport properties in other situations.
This switchable behavior results from a careful design (physical properties and geometry) of the sample structure, which is required to fulfill inequalities between the magnitudes of transport mechanisms.

Our estimates of the relative importance of these various mechanisms can guide the design of innovative, switchable structures, but can also potentially shed light on transport dynamics, e.g., in plants subject to embolism during drought events, or during moisture absorption or drying in wood, soil or concrete.
Future work should help resolve the dynamics in more complex geometries (e.g. three-dimensional, non-regular, etc.).

\section{Acknowledgements}

The authors thank Glenn Swan for technical support.
This work was supported by the National Science Foundation (grants NSF-IGERT DGE-0966045 and NSF GK 12 DGE-1045513), the Air Force Office of Scientific Research (FA9550-21-1-0283) and was performed in part at the Cornell NanoScale Facility, a member of the National Nanotechnology Coordinated Infrastructure (NNCI), which is supported by the National Science Foundation (Grant NNCI-2025233).
The authors also acknowledge the use of various open-source packages from the Python ecosystem for data analysis and visualization, in particular \emph{numpy} and \emph{matplotlib}.

\bibliography{references}

\begin{thebibliography}{37}%
\makeatletter
\providecommand \@ifxundefined [1]{%
 \@ifx{#1\undefined}
}%
\providecommand \@ifnum [1]{%
 \ifnum #1\expandafter \@firstoftwo
 \else \expandafter \@secondoftwo
 \fi
}%
\providecommand \@ifx [1]{%
 \ifx #1\expandafter \@firstoftwo
 \else \expandafter \@secondoftwo
 \fi
}%
\providecommand \natexlab [1]{#1}%
\providecommand \enquote  [1]{``#1''}%
\providecommand \bibnamefont  [1]{#1}%
\providecommand \bibfnamefont [1]{#1}%
\providecommand \citenamefont [1]{#1}%
\providecommand \href@noop [0]{\@secondoftwo}%
\providecommand \href [0]{\begingroup \@sanitize@url \@href}%
\providecommand \@href[1]{\@@startlink{#1}\@@href}%
\providecommand \@@href[1]{\endgroup#1\@@endlink}%
\providecommand \@sanitize@url [0]{\catcode `\\12\catcode `\$12\catcode `\&12\catcode `\#12\catcode `\^12\catcode `\_12\catcode `\%12\relax}%
\providecommand \@@startlink[1]{}%
\providecommand \@@endlink[0]{}%
\providecommand \url  [0]{\begingroup\@sanitize@url \@url }%
\providecommand \@url [1]{\endgroup\@href {#1}{\urlprefix }}%
\providecommand \urlprefix  [0]{URL }%
\providecommand \Eprint [0]{\href }%
\providecommand \doibase [0]{https://doi.org/}%
\providecommand \selectlanguage [0]{\@gobble}%
\providecommand \bibinfo  [0]{\@secondoftwo}%
\providecommand \bibfield  [0]{\@secondoftwo}%
\providecommand \translation [1]{[#1]}%
\providecommand \BibitemOpen [0]{}%
\providecommand \bibitemStop [0]{}%
\providecommand \bibitemNoStop [0]{.\EOS\space}%
\providecommand \EOS [0]{\spacefactor3000\relax}%
\providecommand \BibitemShut  [1]{\csname bibitem#1\endcsname}%
\let\auto@bib@innerbib\@empty
\bibitem [{\citenamefont {Sahimi}(1993)}]{Sahimi1993}%
  \BibitemOpen
  \bibfield  {author} {\bibinfo {author} {\bibfnamefont {M.}~\bibnamefont {Sahimi}},\ }\bibfield  {title} {\bibinfo {title} {Flow phenomena in rocks: From continuum models to fractals, percolation, cellular automata, and simulated annealing},\ }\href {https://doi.org/10.1103/RevModPhys.65.1393} {\bibfield  {journal} {\bibinfo  {journal} {Reviews of Modern Physics}\ }\textbf {\bibinfo {volume} {65}},\ \bibinfo {pages} {1393} (\bibinfo {year} {1993})}\BibitemShut {NoStop}%
\bibitem [{\citenamefont {Huber}(2015)}]{Huber2015}%
  \BibitemOpen
  \bibfield  {author} {\bibinfo {author} {\bibfnamefont {P.}~\bibnamefont {Huber}},\ }\bibfield  {title} {\bibinfo {title} {Soft matter in hard confinement: Phase transition thermodynamics, structure, texture, diffusion and flow in nanoporous media},\ }\href {https://doi.org/10.1088/0953-8984/27/10/103102} {\bibfield  {journal} {\bibinfo  {journal} {Journal of Physics: Condensed Matter}\ }\textbf {\bibinfo {volume} {27}},\ \bibinfo {pages} {103102} (\bibinfo {year} {2015})}\BibitemShut {NoStop}%
\bibitem [{\citenamefont {Bacchin}\ \emph {et~al.}(2021)\citenamefont {Bacchin}, \citenamefont {Leng},\ and\ \citenamefont {Salmon}}]{Bacchin2021}%
  \BibitemOpen
  \bibfield  {author} {\bibinfo {author} {\bibfnamefont {P.}~\bibnamefont {Bacchin}}, \bibinfo {author} {\bibfnamefont {J.}~\bibnamefont {Leng}},\ and\ \bibinfo {author} {\bibfnamefont {J.-B.}\ \bibnamefont {Salmon}},\ }\bibfield  {title} {\bibinfo {title} {Microfluidic {{Evaporation}}, {{Pervaporation}}, and {{Osmosis}}: {{From Passive Pumping}} to {{Solute Concentration}}},\ }\href {https://doi.org/10.1021/acs.chemrev.1c00459} {\bibfield  {journal} {\bibinfo  {journal} {Chemical Reviews}\ ,\ \bibinfo {pages} {acs.chemrev.1c00459}} (\bibinfo {year} {2021})}\BibitemShut {NoStop}%
\bibitem [{\citenamefont {Reising}\ \emph {et~al.}(2017)\citenamefont {Reising}, \citenamefont {Schlabach}, \citenamefont {Baranau}, \citenamefont {Stoeckel},\ and\ \citenamefont {Tallarek}}]{Reising2017}%
  \BibitemOpen
  \bibfield  {author} {\bibinfo {author} {\bibfnamefont {A.~E.}\ \bibnamefont {Reising}}, \bibinfo {author} {\bibfnamefont {S.}~\bibnamefont {Schlabach}}, \bibinfo {author} {\bibfnamefont {V.}~\bibnamefont {Baranau}}, \bibinfo {author} {\bibfnamefont {D.}~\bibnamefont {Stoeckel}},\ and\ \bibinfo {author} {\bibfnamefont {U.}~\bibnamefont {Tallarek}},\ }\bibfield  {title} {\bibinfo {title} {Analysis of packing microstructure and wall effects in a narrow-bore ultrahigh pressure liquid chromatography column using focused ion-beam scanning electron microscopy},\ }\href {https://doi.org/10.1016/j.chroma.2017.07.049} {\bibfield  {journal} {\bibinfo  {journal} {Journal of Chromatography A}\ }\textbf {\bibinfo {volume} {1513}},\ \bibinfo {pages} {172} (\bibinfo {year} {2017})}\BibitemShut {NoStop}%
\bibitem [{\citenamefont {Huber}\ \emph {et~al.}(2018)\citenamefont {Huber}, \citenamefont {Stroock},\ and\ \citenamefont {Koch}}]{Huber2018}%
  \BibitemOpen
  \bibfield  {author} {\bibinfo {author} {\bibfnamefont {E.~J.}\ \bibnamefont {Huber}}, \bibinfo {author} {\bibfnamefont {A.~D.}\ \bibnamefont {Stroock}},\ and\ \bibinfo {author} {\bibfnamefont {D.~L.}\ \bibnamefont {Koch}},\ }\bibfield  {title} {\bibinfo {title} {Modeling the dynamics of remobilized {{CO2}} within the geologic subsurface},\ }\href {https://doi.org/10.1016/j.ijggc.2018.01.020} {\bibfield  {journal} {\bibinfo  {journal} {International Journal of Greenhouse Gas Control}\ }\textbf {\bibinfo {volume} {70}},\ \bibinfo {pages} {128} (\bibinfo {year} {2018})}\BibitemShut {NoStop}%
\bibitem [{\citenamefont {Stuecker}\ \emph {et~al.}(2004)\citenamefont {Stuecker}, \citenamefont {Miller}, \citenamefont {Ferrizz}, \citenamefont {Mudd},\ and\ \citenamefont {Cesarano}}]{Stuecker2004}%
  \BibitemOpen
  \bibfield  {author} {\bibinfo {author} {\bibfnamefont {J.~N.}\ \bibnamefont {Stuecker}}, \bibinfo {author} {\bibfnamefont {J.~E.}\ \bibnamefont {Miller}}, \bibinfo {author} {\bibfnamefont {R.~E.}\ \bibnamefont {Ferrizz}}, \bibinfo {author} {\bibfnamefont {J.~E.}\ \bibnamefont {Mudd}},\ and\ \bibinfo {author} {\bibfnamefont {J.}~\bibnamefont {Cesarano}},\ }\bibfield  {title} {\bibinfo {title} {Advanced {{Support Structures}} for {{Enhanced Catalytic Activity}}},\ }\href {https://doi.org/10.1021/ie030291v} {\bibfield  {journal} {\bibinfo  {journal} {Industrial \& Engineering Chemistry Research}\ }\textbf {\bibinfo {volume} {43}},\ \bibinfo {pages} {51} (\bibinfo {year} {2004})}\BibitemShut {NoStop}%
\bibitem [{\citenamefont {Stroock}\ \emph {et~al.}(2014)\citenamefont {Stroock}, \citenamefont {Pagay}, \citenamefont {Zwieniecki},\ and\ \citenamefont {Holbrook}}]{Stroock2014}%
  \BibitemOpen
  \bibfield  {author} {\bibinfo {author} {\bibfnamefont {A.~D.}\ \bibnamefont {Stroock}}, \bibinfo {author} {\bibfnamefont {V.~V.}\ \bibnamefont {Pagay}}, \bibinfo {author} {\bibfnamefont {M.~A.}\ \bibnamefont {Zwieniecki}},\ and\ \bibinfo {author} {\bibfnamefont {N.~M.}\ \bibnamefont {Holbrook}},\ }\bibfield  {title} {\bibinfo {title} {The {{Physicochemical Hydrodynamics}} of {{Vascular Plants}}},\ }\href {https://doi.org/10.1146/annurev-fluid-010313-141411} {\bibfield  {journal} {\bibinfo  {journal} {Annual Review of Fluid Mechanics}\ }\textbf {\bibinfo {volume} {46}},\ \bibinfo {pages} {615} (\bibinfo {year} {2014})}\BibitemShut {NoStop}%
\bibitem [{\citenamefont {Tyree}\ and\ \citenamefont {Zimmermann}(2002)}]{Tyree2002}%
  \BibitemOpen
  \bibfield  {author} {\bibinfo {author} {\bibfnamefont {M.~T.}\ \bibnamefont {Tyree}}\ and\ \bibinfo {author} {\bibfnamefont {M.~H.}\ \bibnamefont {Zimmermann}},\ }\href@noop {} {\emph {\bibinfo {title} {Xylem Structure and the Ascent of Sap}}},\ \bibinfo {edition} {2nd}\ ed.,\ Springer Series in Wood Science\ (\bibinfo  {publisher} {Springer},\ \bibinfo {address} {Berlin Heidelberg},\ \bibinfo {year} {2002})\BibitemShut {NoStop}%
\bibitem [{\citenamefont {Nobel}(2020)}]{Nobel2020}%
  \BibitemOpen
  \bibfield  {author} {\bibinfo {author} {\bibfnamefont {P.}~\bibnamefont {Nobel}},\ }\href@noop {} {\emph {\bibinfo {title} {Physicochemical and Environmental Plant Physiology}}}\ (\bibinfo  {publisher} {Elsevier},\ \bibinfo {address} {Cambridge},\ \bibinfo {year} {2020})\BibitemShut {NoStop}%
\bibitem [{\citenamefont {Desmarais}\ \emph {et~al.}(2016)\citenamefont {Desmarais}, \citenamefont {Gilani}, \citenamefont {Vontobel}, \citenamefont {Carmeliet},\ and\ \citenamefont {Derome}}]{Desmarais2016}%
  \BibitemOpen
  \bibfield  {author} {\bibinfo {author} {\bibfnamefont {G.}~\bibnamefont {Desmarais}}, \bibinfo {author} {\bibfnamefont {M.~S.}\ \bibnamefont {Gilani}}, \bibinfo {author} {\bibfnamefont {P.}~\bibnamefont {Vontobel}}, \bibinfo {author} {\bibfnamefont {J.}~\bibnamefont {Carmeliet}},\ and\ \bibinfo {author} {\bibfnamefont {D.}~\bibnamefont {Derome}},\ }\bibfield  {title} {\bibinfo {title} {Transport of {{Polar}} and {{Nonpolar Liquids}} in {{Softwood Imaged}} by {{Neutron Radiography}}},\ }\href {https://doi.org/10.1007/s11242-016-0700-4} {\bibfield  {journal} {\bibinfo  {journal} {Transport in Porous Media}\ }\textbf {\bibinfo {volume} {113}},\ \bibinfo {pages} {383} (\bibinfo {year} {2016})}\BibitemShut {NoStop}%
\bibitem [{\citenamefont {Zhou}\ \emph {et~al.}(2018)\citenamefont {Zhou}, \citenamefont {Car{\'e}}, \citenamefont {{Courtier-Murias}}, \citenamefont {Faure}, \citenamefont {Rodts},\ and\ \citenamefont {Coussot}}]{Zhou2018a}%
  \BibitemOpen
  \bibfield  {author} {\bibinfo {author} {\bibfnamefont {M.}~\bibnamefont {Zhou}}, \bibinfo {author} {\bibfnamefont {S.}~\bibnamefont {Car{\'e}}}, \bibinfo {author} {\bibfnamefont {D.}~\bibnamefont {{Courtier-Murias}}}, \bibinfo {author} {\bibfnamefont {P.}~\bibnamefont {Faure}}, \bibinfo {author} {\bibfnamefont {S.}~\bibnamefont {Rodts}},\ and\ \bibinfo {author} {\bibfnamefont {P.}~\bibnamefont {Coussot}},\ }\bibfield  {title} {\bibinfo {title} {Magnetic resonance imaging evidences of the impact of water sorption on hardwood capillary imbibition dynamics},\ }\href {https://doi.org/10.1007/s00226-018-1017-y} {\bibfield  {journal} {\bibinfo  {journal} {Wood Science and Technology}\ }\textbf {\bibinfo {volume} {52}},\ \bibinfo {pages} {929} (\bibinfo {year} {2018})}\BibitemShut {NoStop}%
\bibitem [{\citenamefont {Elustondo}\ \emph {et~al.}(2023)\citenamefont {Elustondo}, \citenamefont {Matan}, \citenamefont {Langrish},\ and\ \citenamefont {Pang}}]{Elustondo2023}%
  \BibitemOpen
  \bibfield  {author} {\bibinfo {author} {\bibfnamefont {D.}~\bibnamefont {Elustondo}}, \bibinfo {author} {\bibfnamefont {N.}~\bibnamefont {Matan}}, \bibinfo {author} {\bibfnamefont {T.}~\bibnamefont {Langrish}},\ and\ \bibinfo {author} {\bibfnamefont {S.}~\bibnamefont {Pang}},\ }\bibfield  {title} {\bibinfo {title} {Advances in wood drying research and development},\ }\href {https://doi.org/10.1080/07373937.2023.2205530} {\bibfield  {journal} {\bibinfo  {journal} {Drying Technology}\ }\textbf {\bibinfo {volume} {41}},\ \bibinfo {pages} {890} (\bibinfo {year} {2023})}\BibitemShut {NoStop}%
\bibitem [{\citenamefont {Wheeler}\ and\ \citenamefont {Stroock}(2008)}]{Wheeler2008}%
  \BibitemOpen
  \bibfield  {author} {\bibinfo {author} {\bibfnamefont {T.~D.}\ \bibnamefont {Wheeler}}\ and\ \bibinfo {author} {\bibfnamefont {A.~D.}\ \bibnamefont {Stroock}},\ }\bibfield  {title} {\bibinfo {title} {The transpiration of water at negative pressures in a synthetic tree},\ }\href {https://doi.org/10.1038/nature07226} {\bibfield  {journal} {\bibinfo  {journal} {Nature}\ }\textbf {\bibinfo {volume} {455}},\ \bibinfo {pages} {208} (\bibinfo {year} {2008})}\BibitemShut {NoStop}%
\bibitem [{\citenamefont {Vincent}\ \emph {et~al.}(2014)\citenamefont {Vincent}, \citenamefont {Sessoms}, \citenamefont {Huber}, \citenamefont {Guioth},\ and\ \citenamefont {Stroock}}]{Vincent2014}%
  \BibitemOpen
  \bibfield  {author} {\bibinfo {author} {\bibfnamefont {O.}~\bibnamefont {Vincent}}, \bibinfo {author} {\bibfnamefont {D.~A.}\ \bibnamefont {Sessoms}}, \bibinfo {author} {\bibfnamefont {E.~J.}\ \bibnamefont {Huber}}, \bibinfo {author} {\bibfnamefont {J.}~\bibnamefont {Guioth}},\ and\ \bibinfo {author} {\bibfnamefont {A.~D.}\ \bibnamefont {Stroock}},\ }\bibfield  {title} {\bibinfo {title} {Drying by {{Cavitation}} and {{Poroelastic Relaxations}} in {{Porous Media}} with {{Macroscopic Pores Connected}} by {{Nanoscale Throats}}},\ }\href {https://doi.org/10.1103/PhysRevLett.113.134501} {\bibfield  {journal} {\bibinfo  {journal} {Physical Review Letters}\ }\textbf {\bibinfo {volume} {113}},\ \bibinfo {pages} {134501} (\bibinfo {year} {2014})}\BibitemShut {NoStop}%
\bibitem [{\citenamefont {Pagay}\ \emph {et~al.}(2014)\citenamefont {Pagay}, \citenamefont {Santiago}, \citenamefont {Sessoms}, \citenamefont {Huber}, \citenamefont {Vincent}, \citenamefont {Pharkya}, \citenamefont {Corso}, \citenamefont {Lakso},\ and\ \citenamefont {Stroock}}]{Pagay2014}%
  \BibitemOpen
  \bibfield  {author} {\bibinfo {author} {\bibfnamefont {V.}~\bibnamefont {Pagay}}, \bibinfo {author} {\bibfnamefont {M.}~\bibnamefont {Santiago}}, \bibinfo {author} {\bibfnamefont {D.~A.}\ \bibnamefont {Sessoms}}, \bibinfo {author} {\bibfnamefont {E.~J.}\ \bibnamefont {Huber}}, \bibinfo {author} {\bibfnamefont {O.}~\bibnamefont {Vincent}}, \bibinfo {author} {\bibfnamefont {A.}~\bibnamefont {Pharkya}}, \bibinfo {author} {\bibfnamefont {T.~N.}\ \bibnamefont {Corso}}, \bibinfo {author} {\bibfnamefont {A.~N.}\ \bibnamefont {Lakso}},\ and\ \bibinfo {author} {\bibfnamefont {A.~D.}\ \bibnamefont {Stroock}},\ }\bibfield  {title} {\bibinfo {title} {A microtensiometer capable of measuring water potentials below -10 {{MPa}}},\ }\href {https://doi.org/10.1039/C4LC00342J} {\bibfield  {journal} {\bibinfo  {journal} {Lab on a Chip}\ }\textbf {\bibinfo {volume} {14}},\ \bibinfo {pages} {2806} (\bibinfo {year} {2014})}\BibitemShut {NoStop}%
\bibitem [{\citenamefont {Vincent}\ \emph {et~al.}(2019)\citenamefont {Vincent}, \citenamefont {Zhang}, \citenamefont {Choi}, \citenamefont {Zhu},\ and\ \citenamefont {Stroock}}]{Vincent2019}%
  \BibitemOpen
  \bibfield  {author} {\bibinfo {author} {\bibfnamefont {O.}~\bibnamefont {Vincent}}, \bibinfo {author} {\bibfnamefont {J.}~\bibnamefont {Zhang}}, \bibinfo {author} {\bibfnamefont {E.}~\bibnamefont {Choi}}, \bibinfo {author} {\bibfnamefont {S.}~\bibnamefont {Zhu}},\ and\ \bibinfo {author} {\bibfnamefont {A.~D.}\ \bibnamefont {Stroock}},\ }\bibfield  {title} {\bibinfo {title} {How {{Solutes Modify}} the {{Thermodynamics}} and {{Dynamics}} of {{Filling}} and {{Emptying}} in {{Extreme Ink-Bottle Pores}}},\ }\href {https://doi.org/10.1021/acs.langmuir.8b03494} {\bibfield  {journal} {\bibinfo  {journal} {Langmuir}\ }\textbf {\bibinfo {volume} {35}},\ \bibinfo {pages} {2934} (\bibinfo {year} {2019})}\BibitemShut {NoStop}%
\bibitem [{\citenamefont {Vincent}\ \emph {et~al.}(2016)\citenamefont {Vincent}, \citenamefont {Szenicer},\ and\ \citenamefont {Stroock}}]{Vincent2016}%
  \BibitemOpen
  \bibfield  {author} {\bibinfo {author} {\bibfnamefont {O.}~\bibnamefont {Vincent}}, \bibinfo {author} {\bibfnamefont {A.}~\bibnamefont {Szenicer}},\ and\ \bibinfo {author} {\bibfnamefont {A.~D.}\ \bibnamefont {Stroock}},\ }\bibfield  {title} {\bibinfo {title} {Capillarity-driven flows at the continuum limit},\ }\href {https://doi.org/10.1039/C6SM00733C} {\bibfield  {journal} {\bibinfo  {journal} {Soft Matter}\ }\textbf {\bibinfo {volume} {12}},\ \bibinfo {pages} {6656} (\bibinfo {year} {2016})}\BibitemShut {NoStop}%
\bibitem [{\citenamefont {Chen}\ \emph {et~al.}(2016)\citenamefont {Chen}, \citenamefont {Sessoms}, \citenamefont {Sherman}, \citenamefont {Choi}, \citenamefont {Vincent},\ and\ \citenamefont {Stroock}}]{Chen2016a}%
  \BibitemOpen
  \bibfield  {author} {\bibinfo {author} {\bibfnamefont {I.-T.}\ \bibnamefont {Chen}}, \bibinfo {author} {\bibfnamefont {D.~A.}\ \bibnamefont {Sessoms}}, \bibinfo {author} {\bibfnamefont {Z.}~\bibnamefont {Sherman}}, \bibinfo {author} {\bibfnamefont {E.}~\bibnamefont {Choi}}, \bibinfo {author} {\bibfnamefont {O.}~\bibnamefont {Vincent}},\ and\ \bibinfo {author} {\bibfnamefont {A.~D.}\ \bibnamefont {Stroock}},\ }\bibfield  {title} {\bibinfo {title} {Stability {{Limit}} of {{Water}} by {{Metastable Vapor}}--{{Liquid Equilibrium}} with {{Nanoporous Silicon Membranes}}},\ }\href {https://doi.org/10.1021/acs.jpcb.6b01618} {\bibfield  {journal} {\bibinfo  {journal} {The Journal of Physical Chemistry B}\ }\textbf {\bibinfo {volume} {120}},\ \bibinfo {pages} {5209} (\bibinfo {year} {2016})}\BibitemShut {NoStop}%
\bibitem [{\citenamefont {Shi}\ \emph {et~al.}(2020)\citenamefont {Shi}, \citenamefont {Dalrymple}, \citenamefont {McKenny}, \citenamefont {Morrow}, \citenamefont {Rashed}, \citenamefont {Surinach},\ and\ \citenamefont {Boreyko}}]{Shi2020}%
  \BibitemOpen
  \bibfield  {author} {\bibinfo {author} {\bibfnamefont {W.}~\bibnamefont {Shi}}, \bibinfo {author} {\bibfnamefont {R.~M.}\ \bibnamefont {Dalrymple}}, \bibinfo {author} {\bibfnamefont {C.~J.}\ \bibnamefont {McKenny}}, \bibinfo {author} {\bibfnamefont {D.~S.}\ \bibnamefont {Morrow}}, \bibinfo {author} {\bibfnamefont {Z.~T.}\ \bibnamefont {Rashed}}, \bibinfo {author} {\bibfnamefont {D.~A.}\ \bibnamefont {Surinach}},\ and\ \bibinfo {author} {\bibfnamefont {J.~B.}\ \bibnamefont {Boreyko}},\ }\bibfield  {title} {\bibinfo {title} {Passive water ascent in a tall, scalable synthetic tree},\ }\href {https://doi.org/10.1038/s41598-019-57109-z} {\bibfield  {journal} {\bibinfo  {journal} {Scientific Reports}\ }\textbf {\bibinfo {volume} {10}},\ \bibinfo {pages} {1} (\bibinfo {year} {2020})}\BibitemShut {NoStop}%
\bibitem [{\citenamefont {Wang}\ \emph {et~al.}(2020)\citenamefont {Wang}, \citenamefont {Lee}, \citenamefont {Werber},\ and\ \citenamefont {Elimelech}}]{Wang2020}%
  \BibitemOpen
  \bibfield  {author} {\bibinfo {author} {\bibfnamefont {Y.}~\bibnamefont {Wang}}, \bibinfo {author} {\bibfnamefont {J.}~\bibnamefont {Lee}}, \bibinfo {author} {\bibfnamefont {J.~R.}\ \bibnamefont {Werber}},\ and\ \bibinfo {author} {\bibfnamefont {M.}~\bibnamefont {Elimelech}},\ }\bibfield  {title} {\bibinfo {title} {Capillary-driven desalination in a synthetic mangrove},\ }\href {https://doi.org/10.1126/sciadv.aax5253} {\bibfield  {journal} {\bibinfo  {journal} {Science Advances}\ }\textbf {\bibinfo {volume} {6}},\ \bibinfo {pages} {eaax5253} (\bibinfo {year} {2020})}\BibitemShut {NoStop}%
\bibitem [{\citenamefont {Vincent}\ \emph {et~al.}(2017)\citenamefont {Vincent}, \citenamefont {Marguet},\ and\ \citenamefont {Stroock}}]{Vincent2017a}%
  \BibitemOpen
  \bibfield  {author} {\bibinfo {author} {\bibfnamefont {O.}~\bibnamefont {Vincent}}, \bibinfo {author} {\bibfnamefont {B.}~\bibnamefont {Marguet}},\ and\ \bibinfo {author} {\bibfnamefont {A.~D.}\ \bibnamefont {Stroock}},\ }\bibfield  {title} {\bibinfo {title} {Imbibition {{Triggered}} by {{Capillary Condensation}} in {{Nanopores}}},\ }\href {https://doi.org/10.1021/acs.langmuir.6b04534} {\bibfield  {journal} {\bibinfo  {journal} {Langmuir}\ }\textbf {\bibinfo {volume} {33}},\ \bibinfo {pages} {1655} (\bibinfo {year} {2017})}\BibitemShut {NoStop}%
\bibitem [{Note1()}]{Note1}%
  \BibitemOpen
  \bibinfo {note} {Note that appropriate equipment and protection must be used when handling fluids at high pressures.}\BibitemShut {Stop}%
\bibitem [{\citenamefont {Wagner}\ and\ \citenamefont {Pru{\ss}}(2002)}]{Wagner2002}%
  \BibitemOpen
  \bibfield  {author} {\bibinfo {author} {\bibfnamefont {W.}~\bibnamefont {Wagner}}\ and\ \bibinfo {author} {\bibfnamefont {A.}~\bibnamefont {Pru{\ss}}},\ }\bibfield  {title} {\bibinfo {title} {The {{IAPWS Formulation}} 1995 for the {{Thermodynamic Properties}} of {{Ordinary Water Substance}} for {{General}} and {{Scientific Use}}},\ }\href {https://doi.org/10.1063/1.1461829} {\bibfield  {journal} {\bibinfo  {journal} {Journal of Physical and Chemical Reference Data}\ }\textbf {\bibinfo {volume} {31}},\ \bibinfo {pages} {387} (\bibinfo {year} {2002})}\BibitemShut {NoStop}%
\bibitem [{\citenamefont {Huber}\ \emph {et~al.}(2009)\citenamefont {Huber}, \citenamefont {Perkins}, \citenamefont {Laesecke}, \citenamefont {Friend}, \citenamefont {Sengers}, \citenamefont {Assael}, \citenamefont {Metaxa}, \citenamefont {Vogel}, \citenamefont {Mare{\v s}},\ and\ \citenamefont {Miyagawa}}]{Huber2009}%
  \BibitemOpen
  \bibfield  {author} {\bibinfo {author} {\bibfnamefont {M.~L.}\ \bibnamefont {Huber}}, \bibinfo {author} {\bibfnamefont {R.~A.}\ \bibnamefont {Perkins}}, \bibinfo {author} {\bibfnamefont {A.}~\bibnamefont {Laesecke}}, \bibinfo {author} {\bibfnamefont {D.~G.}\ \bibnamefont {Friend}}, \bibinfo {author} {\bibfnamefont {J.~V.}\ \bibnamefont {Sengers}}, \bibinfo {author} {\bibfnamefont {M.~J.}\ \bibnamefont {Assael}}, \bibinfo {author} {\bibfnamefont {I.~N.}\ \bibnamefont {Metaxa}}, \bibinfo {author} {\bibfnamefont {E.}~\bibnamefont {Vogel}}, \bibinfo {author} {\bibfnamefont {R.}~\bibnamefont {Mare{\v s}}},\ and\ \bibinfo {author} {\bibfnamefont {K.}~\bibnamefont {Miyagawa}},\ }\bibfield  {title} {\bibinfo {title} {New {{International Formulation}} for the {{Viscosity}} of {{H2O}}},\ }\href {https://doi.org/10.1063/1.3088050} {\bibfield  {journal} {\bibinfo  {journal} {Journal of Physical and Chemical Reference Data}\ }\textbf {\bibinfo {volume} {38}},\ \bibinfo {pages} {101} (\bibinfo {year} {2009})}\BibitemShut {NoStop}%
\bibitem [{\citenamefont {Massman}(1998)}]{Massman1998}%
  \BibitemOpen
  \bibfield  {author} {\bibinfo {author} {\bibfnamefont {W.~J.}\ \bibnamefont {Massman}},\ }\bibfield  {title} {\bibinfo {title} {A review of the molecular diffusivities of {{H2O}}, {{CO2}}, {{CH4}}, {{CO}}, {{O3}}, {{SO2}}, {{NH3}}, {{N2O}}, {{NO}}, and {{NO2}} in air, {{O2}} and {{N2}} near {{STP}}},\ }\href {https://doi.org/10.1016/S1352-2310(97)00391-9} {\bibfield  {journal} {\bibinfo  {journal} {Atmospheric Environment}\ }\textbf {\bibinfo {volume} {32}},\ \bibinfo {pages} {1111} (\bibinfo {year} {1998})}\BibitemShut {NoStop}%
\bibitem [{\citenamefont {Vincent}(2022)}]{Vincent2022}%
  \BibitemOpen
  \bibfield  {author} {\bibinfo {author} {\bibfnamefont {O.}~\bibnamefont {Vincent}},\ }\bibfield  {title} {\bibinfo {title} {Chapter 4. {{Negative Pressure}} and {{Cavitation Dynamics}} in {{Plant-like Structures}}},\ }in\ \href {https://doi.org/10.1039/9781839161162-00119} {\emph {\bibinfo {booktitle} {Soft {{Matter}} in {{Plants}}: {{From Biophysics}} to {{Biomimetics}}}}},\ \bibinfo {series and number} {Soft {{Matter Series}}},\ \bibinfo {editor} {edited by\ \bibinfo {editor} {\bibfnamefont {K.}~\bibnamefont {Jensen}}\ and\ \bibinfo {editor} {\bibfnamefont {Y.}~\bibnamefont {Forterre}}}\ (\bibinfo  {publisher} {Royal Society of Chemistry},\ \bibinfo {address} {Cambridge},\ \bibinfo {year} {2022})\ pp.\ \bibinfo {pages} {119--164}\BibitemShut {NoStop}%
\bibitem [{\citenamefont {Pingulkar}\ \emph {et~al.}(2024)\citenamefont {Pingulkar}, \citenamefont {Mar{\'e}chal},\ and\ \citenamefont {Salmon}}]{Pingulkar2024}%
  \BibitemOpen
  \bibfield  {author} {\bibinfo {author} {\bibfnamefont {H.}~\bibnamefont {Pingulkar}}, \bibinfo {author} {\bibfnamefont {S.}~\bibnamefont {Mar{\'e}chal}},\ and\ \bibinfo {author} {\bibfnamefont {J.-B.}\ \bibnamefont {Salmon}},\ }\bibfield  {title} {\bibinfo {title} {Directional drying of a colloidal dispersion: Quantitative description with water potential measurements using water clusters in a poly(dimethylsiloxane) microfluidic chip},\ }\href {https://doi.org/10.1039/D3SM01512B} {\bibfield  {journal} {\bibinfo  {journal} {Soft Matter}\ ,\ \bibinfo {pages} {10.1039.D3SM01512B}} (\bibinfo {year} {2024})}\BibitemShut {NoStop}%
\bibitem [{\citenamefont {Choy}(2016)}]{Choy2016}%
  \BibitemOpen
  \bibfield  {author} {\bibinfo {author} {\bibfnamefont {T.~C.}\ \bibnamefont {Choy}},\ }\href@noop {} {\emph {\bibinfo {title} {Effective Medium Theory: Principles and Applications}}},\ \bibinfo {edition} {second edition}\ ed.,\ \bibinfo {series} {International Series of Monographs on Physics}\ No.\ \bibinfo {number} {165}\ (\bibinfo  {publisher} {Oxford University Press},\ \bibinfo {address} {Oxford},\ \bibinfo {year} {2016})\BibitemShut {NoStop}%
\bibitem [{\citenamefont {Renard}\ and\ \citenamefont {{de Marsily}}(1997)}]{Renard1997}%
  \BibitemOpen
  \bibfield  {author} {\bibinfo {author} {\bibfnamefont {{\relax Ph}.}~\bibnamefont {Renard}}\ and\ \bibinfo {author} {\bibfnamefont {G.}~\bibnamefont {{de Marsily}}},\ }\bibfield  {title} {\bibinfo {title} {Calculating equivalent permeability: A review},\ }\href {https://doi.org/10.1016/S0309-1708(96)00050-4} {\bibfield  {journal} {\bibinfo  {journal} {Advances in Water Resources}\ }\textbf {\bibinfo {volume} {20}},\ \bibinfo {pages} {253} (\bibinfo {year} {1997})}\BibitemShut {NoStop}%
\bibitem [{\citenamefont {Huber}(2017)}]{Huber2017}%
  \BibitemOpen
  \bibfield  {author} {\bibinfo {author} {\bibfnamefont {E.~J.}\ \bibnamefont {Huber}},\ }\emph {\bibinfo {title} {Modeling the {{Dynamics}} of {{Carbon Sequestration}}: {{Injection Strategies}}, {{Remobilization}}, and {{Cavitation}}}},\ \href@noop {} {Ph.D. thesis},\ \bibinfo  {school} {Cornell University} (\bibinfo {year} {2017})\BibitemShut {NoStop}%
\bibitem [{\citenamefont {Garnett}(1904)}]{Garnett1904}%
  \BibitemOpen
  \bibfield  {author} {\bibinfo {author} {\bibfnamefont {J.~C.~M.}\ \bibnamefont {Garnett}},\ }\bibfield  {title} {\bibinfo {title} {{{XII}}. {{Colours}} in metal glasses and in metallic films},\ }\href {https://doi.org/10.1098/rsta.1904.0024} {\bibfield  {journal} {\bibinfo  {journal} {Philosophical Transactions of the Royal Society of London. Series A, Containing Papers of a Mathematical or Physical Character}\ }\textbf {\bibinfo {volume} {203}},\ \bibinfo {pages} {385} (\bibinfo {year} {1904})}\BibitemShut {NoStop}%
\bibitem [{\citenamefont {Zimmerman}(1996)}]{Zimmerman1996}%
  \BibitemOpen
  \bibfield  {author} {\bibinfo {author} {\bibfnamefont {R.~W.}\ \bibnamefont {Zimmerman}},\ }\bibfield  {title} {\bibinfo {title} {Effective conductivity of a two-dimensional medium containing elliptical inhomogeneities},\ }\href@noop {} {\bibfield  {journal} {\bibinfo  {journal} {Proceedings of the Royal Society of London. Series A: Mathematical, Physical and Engineering Sciences}\ }\textbf {\bibinfo {volume} {452}},\ \bibinfo {pages} {1713} (\bibinfo {year} {1996})}\BibitemShut {NoStop}%
\bibitem [{\citenamefont {Lucas}(1918)}]{Lucas1918}%
  \BibitemOpen
  \bibfield  {author} {\bibinfo {author} {\bibfnamefont {R.}~\bibnamefont {Lucas}},\ }\bibfield  {title} {\bibinfo {title} {{Ueber das Zeitgesetz des kapillaren Aufstiegs von Fl{\"u}ssigkeiten}},\ }\href {https://doi.org/10.1007/BF01461107} {\bibfield  {journal} {\bibinfo  {journal} {Kolloid-Zeitschrift}\ }\textbf {\bibinfo {volume} {23}},\ \bibinfo {pages} {15} (\bibinfo {year} {1918})}\BibitemShut {NoStop}%
\bibitem [{\citenamefont {Washburn}(1921)}]{Washburn1921a}%
  \BibitemOpen
  \bibfield  {author} {\bibinfo {author} {\bibfnamefont {E.~W.}\ \bibnamefont {Washburn}},\ }\bibfield  {title} {\bibinfo {title} {The {{Dynamics}} of {{Capillary Flow}}},\ }\href {https://doi.org/10.1103/PhysRev.17.273} {\bibfield  {journal} {\bibinfo  {journal} {Physical Review}\ }\textbf {\bibinfo {volume} {17}},\ \bibinfo {pages} {273} (\bibinfo {year} {1921})}\BibitemShut {NoStop}%
\bibitem [{\citenamefont {Caupin}\ and\ \citenamefont {Stroock}(2013)}]{Caupin2013}%
  \BibitemOpen
  \bibfield  {author} {\bibinfo {author} {\bibfnamefont {F.}~\bibnamefont {Caupin}}\ and\ \bibinfo {author} {\bibfnamefont {A.~D.}\ \bibnamefont {Stroock}},\ }\bibfield  {title} {\bibinfo {title} {The {{Stability Limit}} and other {{Open Questions}} on {{Water}} at {{Negative Pressure}}},\ }in\ \href {https://doi.org/10.1002/9781118540350.ch3} {\emph {\bibinfo {booktitle} {Advances in {{Chemical Physics}}}}},\ \bibinfo {editor} {edited by\ \bibinfo {editor} {\bibfnamefont {H.~E.}\ \bibnamefont {Stanley}}}\ (\bibinfo  {publisher} {John Wiley \& Sons, Inc.},\ \bibinfo {address} {Hoboken, NJ, USA},\ \bibinfo {year} {2013})\ pp.\ \bibinfo {pages} {51--80}\BibitemShut {NoStop}%
\bibitem [{\citenamefont {Cochard}(2006)}]{Cochard2006}%
  \BibitemOpen
  \bibfield  {author} {\bibinfo {author} {\bibfnamefont {H.}~\bibnamefont {Cochard}},\ }\bibfield  {title} {\bibinfo {title} {Cavitation in trees},\ }\href {https://doi.org/10.1016/j.crhy.2006.10.012} {\bibfield  {journal} {\bibinfo  {journal} {Comptes Rendus Physique}\ }\textbf {\bibinfo {volume} {7}},\ \bibinfo {pages} {1018} (\bibinfo {year} {2006})}\BibitemShut {NoStop}%
\bibitem [{\citenamefont {Caupin}\ \emph {et~al.}(2008)\citenamefont {Caupin}, \citenamefont {Herbert}, \citenamefont {Balibar},\ and\ \citenamefont {Cole}}]{Caupin2008}%
  \BibitemOpen
  \bibfield  {author} {\bibinfo {author} {\bibfnamefont {F.}~\bibnamefont {Caupin}}, \bibinfo {author} {\bibfnamefont {E.}~\bibnamefont {Herbert}}, \bibinfo {author} {\bibfnamefont {S.}~\bibnamefont {Balibar}},\ and\ \bibinfo {author} {\bibfnamefont {M.~W.}\ \bibnamefont {Cole}},\ }\bibfield  {title} {\bibinfo {title} {Comment on `{{Nanoscale}} water capillary bridges under deeply negative pressure' [{{Chem}}. {{Phys}}. {{Lett}}. 451 (2008) 88]},\ }\href {https://doi.org/10.1016/j.cplett.2008.08.047} {\bibfield  {journal} {\bibinfo  {journal} {Chemical Physics Letters}\ }\textbf {\bibinfo {volume} {463}},\ \bibinfo {pages} {283} (\bibinfo {year} {2008})}\BibitemShut {NoStop}%
\end{thebibliography}%

\newpage

\onecolumngrid

\appendix

\section{Sample pictures \label{sec:AppendixSamples}}

\begin{figure}[]
	\includegraphics[scale=0.8]{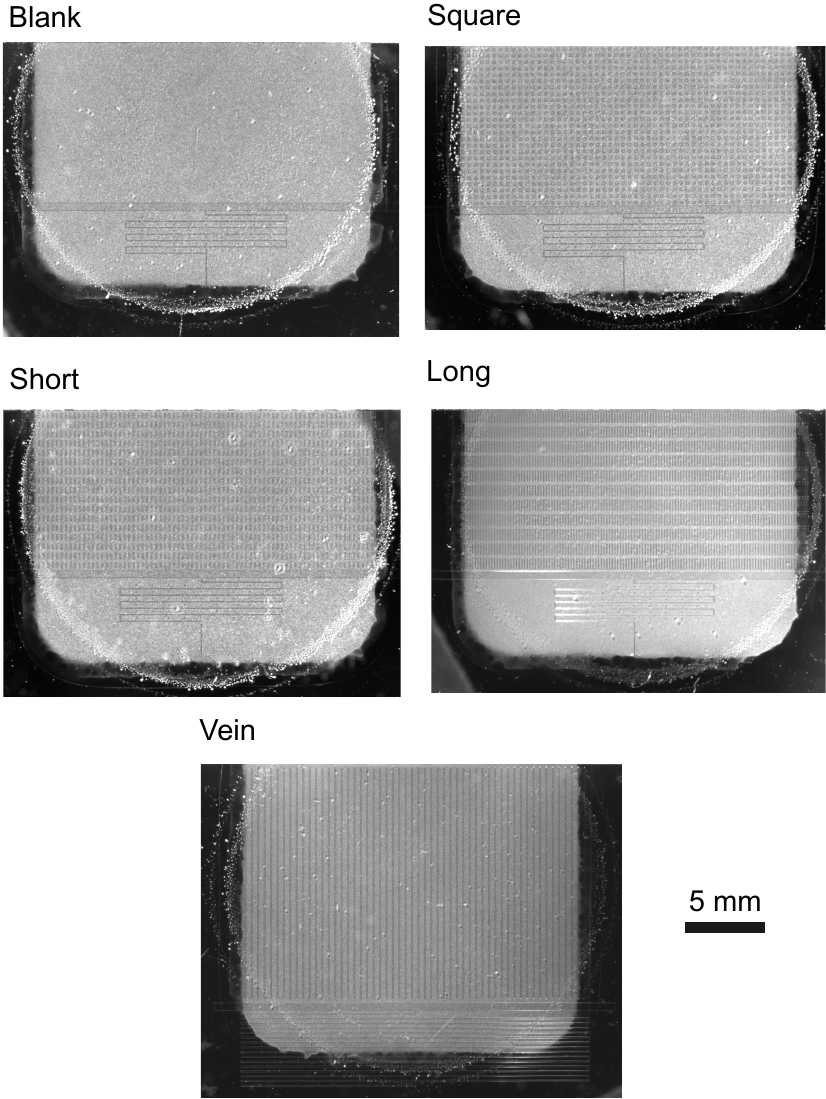}
		\caption{\small
		Micrographs of all samples used in this study. The scale bar serves for all images.
		}
	\label{fig:SamplePictures}
\end{figure}

For the present study, we have used five different samples, the properties of which being listed in Table \ref{tab:Samples}. In Fig. \ref{fig:SamplePictures}, we also provide actual micrographs of all samples. The micrographs for the square and long samples are identical to those in Fig. \ref{fig:Samples}b.

\section{High pressure experiment \label{sec:AppendixPBomb}}

\begin{figure}[]
	\includegraphics[scale=1]{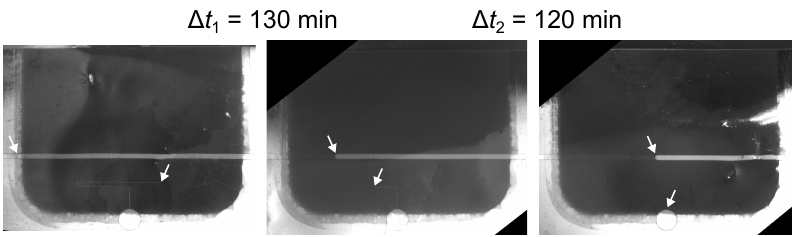}
		\caption{\small
		Example of high pressure-driven flow, here with the \emph{Short} sample, at 2500 psi ($\simeq 17$ MPa) driving force (contrast has been enhanced for clarity).
		The sample was taken out of the pressurized vessel every two hours approximately and its filling state was measured from micrographs. Typically, the meniscus split between the large channel and the serpentine channel, so that we had to track several liquid-vapor interfaces to measure the filled volume variations (white arrows).
		}
	\label{fig:ForcedFilling}
\end{figure}

In sections \ref{sec:MethodsPBomb} and \ref{sec:ResultsPBomb} we have described high pressure flow experiments, from which the permeability of the composite structures can be estimated. In Fig. \ref{fig:ForcedFilling}, we have provided an example of a partial image sequence obtained in one of these experiments.


\section{Maxwell-Garnett approach for predicting $\kappa / \kappa_0$ \label{sec:AppendixMaxwell}}

In order to predict the permeability, $\kappa$ of two-dimensional composite structures containing arrays of rectangular inclusions (axial aspect ratio, $\xi = \ell / w$, covered areal fraction, $\gamma$) of infinite permeability embedded in a nanoporous matrix, we adapt a calculation based on Maxwell-Garnett effective medium approach \cite{Garnett1904,Choy2016}, made by Zimmerman \cite{Zimmerman1996} for ellipses.
Our derivation (see also \cite{Huber2017}) assumes that ellipses and rectangles with the same geometrical properties ($\xi$, $\gamma$) should have similar effects on effective permeability.

Zimmerman considers a homogeneous matrix (permeability, $\kappa_0$), with an unperturbed flow of constant velocity, $V_0$, aligned with an axis, $X$, created by an imposed pressure gradient.
Zimmermann then evaluates the perturbation induced by circular region $\mathcal{A}$ of radius, $R$, added within the matrix.
This region contains $N$ ellipses of permeability $\kappa_1$, randomly distributed and randomly oriented (angle, $\theta$, with respect to the flow direction).
By summing the contribution of all individual ellipses, Zimmerman shows that the perturbed velocity field at some large distance outside of $\mathcal{A}$ follows
\begin{equation}
	\bar{V} = V_0
	\left[
		1 -
		\frac{
			\gamma ( 1 + \alpha ) (1 - r)
		}
		{
			2 (1 + \alpha r) (\alpha + r)
		}
		\left(\frac{R}{\bar{Z}}\right)^2
		\langle
			[
				(\alpha + r) \cos \theta - \mathrm{i} (1 + \alpha r) \sin \theta
			]
			\exp(\mathrm{i} \theta)
		\rangle
	\right]
	\label{eq:FlowFieldZimmermann}
\end{equation}
where $\bar{V} = V_X + \mathrm{i} V_Y$ is the complex velocity field,
$\bar{Z} = X + \mathrm{i}Y$ is the complex position with respect to the center of $\mathcal{A}$,
$r = \kappa_1 / \kappa_0$ is the permeability contrast between the ellipses and the matrix,
$\alpha = 1 / \xi = w / \ell$ is the lateral aspect ratio of the ellipses, and $\langle \rangle$ is an average across all orientations.
Note that compared to Equation (2.5) in Zimmerman's article, we have already replaced $N \ell^2 \alpha$ by $\gamma R^2$, using the fact that the areal ratio is $\gamma = \pi \ell w / (\pi R^2)$ and $w = \alpha \ell$.

In our case all microchannels are aligned, with an angle $\theta = 0$ with respect to the flow.
As a result, the average $\langle \cdots \rangle$ evaluates to $\alpha + r$ and Equation (\ref{eq:FlowFieldZimmermann}) becomes
\begin{equation}
	\bar{V} = V_0
	\left[
		1 -
		\frac{
			\gamma ( 1 + \alpha ) (1 - r)
		}
		{
			2 (1 + \alpha r)
		}
		\left(\frac{R}{\bar{Z}}\right)^2
	\right]
	\label{eq:FlowFieldErik}
\end{equation}

Maxwell Garnett's effective medium approach consists in equating this flow field with that created by a single, circular void with the same radius, $R$, behaving as a homogeneous medium with effective permeability, $\kappa$ \cite{Zimmerman1996}:
\begin{equation}
	\bar{V}
	=
	V_0
	\left[
		1 -
		\frac{
			\kappa_0 - \kappa
		}
		{
			\kappa_0 + \kappa
		}
		\left(\frac{R}{\bar{Z}}\right)^2
	\right]
	\label{eq:FlowFieldCircularArea}
\end{equation}
Equating Equations (\ref{eq:FlowFieldErik}) and (\ref{eq:FlowFieldCircularArea}), one can show that the effective permeability follows \cite{Huber2017}
\begin{equation}
	\frac{\kappa}{\kappa_0}
	=
	\frac{
		1 - \beta \gamma
	}
	{
		1 + \beta \gamma
	}
	\label{eq:EffectivePermeabilityErik}
\end{equation}
with
\begin{equation}
	\beta
	=
	\frac{
		(1 + \alpha)(1 - r)
	}
	{
		2 ( 1 + \alpha r)
	}
	\label{eq:GFactorErikGeneral}
\end{equation}
or, if we consider infinitely conducting inclusions ($r \rightarrow \infty$), $\beta = - (1 + 1/\alpha) / 2$; in other words, since $\alpha = 1 / \xi$,
\begin{equation}
	\beta_\infty
	=
	- \frac{1}{2}
	(1 + \xi)
	\label{eq:GFactorErikInfty}
\end{equation}
Combining Equations (\ref{eq:EffectivePermeabilityErik}) and (\ref{eq:GFactorErikInfty}), we finally obtain
\begin{equation}
	\left(
		\frac{\kappa}{\kappa_0}
	\right)_\infty
	=
	\frac{
		2 + (1 + \xi) \gamma
	}
	{
		2 - (1 + \xi) \gamma
	}
	\label{eq:EffectivePermeabilityFinal}
\end{equation}


\begin{figure}[]
	\includegraphics[scale=1]{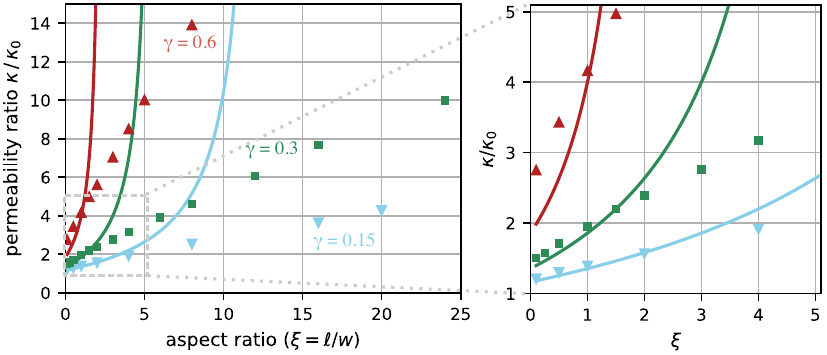}
		\caption{\small
		Comparison between $\kappa / \kappa_0$ obtained from numerical simulations (symbols) and from an analytical, Maxwell Garnett-like approach (lines, Equation (\ref{eq:EffectivePermeabilityFinal})).
		Data from simulations and color-coding are the same as in Figure \ref{fig:Simulations}b from the main text.
		The inset on the right shows a magnification of the data for small aspect ratios, as indicated by the gray rectangle in the Figure.
		Note that $\xi < 1$ corresponds to microchannels with their longest axis perpendicular to the direction of the flow.
		}
	\label{fig:SimulationsMaxwell}
\end{figure}

Figure \ref{fig:SimulationsMaxwell} compares predictions from Equation (\ref{eq:EffectivePermeabilityFinal}) and the results of our numerical simulations (see main text, Figure \ref{fig:Simulations}).
As can be seen in Figure \ref{fig:SimulationsMaxwell}, Maxwell Garnett's approach performs better at low areal fractions; this is expected since this approach assumes low densities of inclusions (ellipses), in order to neglect interactions between neighboring inclusions.

\section{Unit cell filling time \label{sec:AppendixFillingTime}}

We consider a unit cell at the edge of a sample, and we assume that the microchannel within that unit cell has infinite permeability, thus redistributing efficiently water molecules in the whole unit cell instantly.
We also assume that the process limiting the uptake of water by the nanoporous medium in the whole unit cell is the transfer of water between the edge and the microchannel, over a distance $\simeq d$ (with a cross-section $\mathcal{S} = H_\mathrm{p} (w + 2d)$).
Since this nanoporous edge limiting the flow consists in just the nanoporous layer, it has the same permeability as the \emph{Blank} sample, i.e., $\kappa_0$.
Similarly to the case of imbibition described in the main text, the driving force for liquid flow in that nanoporous edge is $\Delta P = \Psi(p) - \Delta P_\mathrm{c}$. Thus, from Darcy's law (Equations 2 and 13 in manuscript), the corresponding mass flow rate $Q = \mathrm{d}m / \mathrm{d}t$ [kg/s] is
\begin{equation}
	Q = \rho \mathcal{S} \kappa_0 \frac{\Delta P}{d}.
	\label{eq:DarcySI}
\end{equation}
The total mass of water required to fill the nanoporous volume contained in the unit cell is $m = \rho \phi H_\mathrm{p} (w + 2d) (\ell + 2d)$; as a result the typical filling time is $\tau \sim m / Q$, or from Equation (\ref{eq:DarcySI}):
\begin{equation}
	\tau \sim \frac{\phi d L_\mathrm{cell}}{\kappa \Delta P}
	\label{eq:FillingTime}
\end{equation}
where $L_\mathrm{cell} = \ell + 2d$ is the total length of the unit cell. One can recognize the Lucas-Washburn "velocity" $\omega_0 = 2 \kappa_0 \Delta P / \phi \, \mathrm{[m^2/s]}$, so that
\begin{equation}
	\tau \sim \frac{2 d L_\mathrm{cell}}{\omega_0}.
	\label{eq:FillingTime2}
\end{equation}

\section{Supplementary Video \label{sec:AppendixVideo}}

The supplementary video shows the dynamics of imbibition of the \emph{Long} sample, under vacuum conditions (imbibition triggered by the capillary condensation of pure water vapor). The video is accelerated approximately 200 times.

\end{document}